\documentclass[11pt,fleqn,thmsa]{article}
\usepackage{fullpage}
\usepackage[ansinew]{inputenc}
\usepackage{verbatim,tikz}
\usepackage{graphicx}
\usepackage{tabularx}
\usepackage{array}
\usepackage{delarray}
\usepackage{dcolumn}
\usepackage{float}
\usepackage{amsmath}
\usepackage{psfrag}
\usepackage{booktabs}
\usepackage{multirow,hhline}
\usepackage{amstext}
\usepackage{amssymb}
\usepackage{fancyhdr}
\usepackage{setspace}
\usepackage{amsthm}
\usepackage{dsfont}
\usepackage{hyperref}
\usepackage{url}
\usepackage{makeidx}
\usepackage{lipsum}
\usepackage{caption}
\usepackage{subcaption}
\usepackage{xr}
\usepackage{adjustbox}
\usepackage[top=1in, bottom=1in, left=0.8in, right=0.8in]{geometry}
\usepackage{datetime}
\newdateformat{monthyeardate}{%
	\monthname[\THEMONTH] \THEYEAR}

\RequirePackage[round,authoryear,longnamesfirst]{natbib}
\hypersetup{citecolor=true,colorlinks=true,allcolors=blue}
\renewcommand{\baselinestretch}{1.0}
\allowdisplaybreaks

\newcommand{\be}{\begin{eqnarray*}}
	\newcommand{\ee}{\end{eqnarray*}}

\newcommand{\avg}{\frac{1}{n}\sum_{i=1}^n}

\newcommand{\usum}{\frac{1}{n(n-1)} \sum_{i \neq j} }
\newcommand{\usumt}{ \frac{1}{n(n-1)(n-2)} \sum_{i \neq j \neq k} }

\newtheorem{remark}{Remark}[section]
\newtheorem{theorem}{Theorem}[section]

\newtheorem{ass}{Assumption}
\newtheorem{rem}{Remark}

\newcommand{\Supp}{\text{Supp}}
\newcommand{\indep}{\perp\!\!\!\perp}

\newcommand{\N}{\mathcal{N}}
\newcommand{\eps}{\varepsilon}
\renewcommand{\epsilon}{\varepsilon}

\usepackage{xr}

\begin{document}
\title{ \large Informational Content of  Factor Structures in \\  Simultaneous  Binary Response Models\footnote{First version: October 2019. We thank the Editor, Serena Ng, two anonymous referees, Arthur Lewbel, and seminar participants at Arizona State University, Emory, Michigan State, Shanghai University of Finance and Economics, University of Arizona, as well as conference participants at the 2015 SEA meetings for helpful comments. We also thank Zhangchi Ma and Qingsong Yao for excellent research assistance. Zhang acknowledges the financial support from Singapore Ministry of Education Tier 2 grant under grant MOE2018-T2-2-169 and the Lee Kong Chian fellowship.}}
{ \author{ \normalsize Shakeeb Khan \\
		%\small Department of Economics\\[-0.8ex]
		\small Boston College 
		%\small \texttt{shakeebk@duke.edu}\\
		\and \normalsize Arnaud Maurel \\
		%\small Department of Economics\\[-0.8ex]
		\small Duke University, NBER and IZA 
		% \small \texttt{mponomar@uwo.ca}
		\and
		\normalsize Yichong Zhang\\
		%\small Department of Economics \\[-.8ex]
		\small Singapore Management  University 
		%\small \texttt{tamer@northwestern.edu}
}}
\date{ \small \monthyeardate\today \\ }

%\begin{document}
\bibliographystyle{econometrica}

\maketitle

%\end{document}	

%\bibliographystyle{econometrica}

%\bibliographstyle{agsm}

%\maketitle

%\bibliographystyle{econometrica}

%%AM: NEED TO UPDATE THE ABSTRACT 

{%\singlespace
	\begin{abstract}
		We study the informational content of factor structures in discrete triangular systems. Factor structures have been employed in a  variety of settings in cross sectional and panel data models, and in this paper we formally quantify their identifying power in a bivariate system often employed in the treatment effects literature. Our main findings are that imposing a factor structure yields point identification of parameters of interest, such as the coefficient associated with the endogenous regressor in the outcome equation, under weaker assumptions than usually required in these models. %Our main findings are that under the factor structures often imposed in the literature,  point identification of parameters of interest, such as both the  treatment effect and the factor load, is attainable under weaker assumptions than usually required in these systems 
		In particular, we show that a ``non-standard'' exclusion restriction that requires an explanatory variable in the outcome equation to be excluded from the treatment equation is no longer necessary for identification, even in cases where all of the regressors from the outcome equation are discrete.
		%Furthermore, we show that support conditions of included covariates in the outcome equation can be substantially weakened, resulting in settings where identification is not based on 
		%(\textbf{Yichong: VY's estimator is also regular in the sense that it is root-n consistent and asymptotically normal. It is not the case that the parameter \textit{becomes} regular with the aid of the factor structure.})
		% Under such settings, we propose a rank estimator for both the factor loading and the causal effect parameter that are root-$n$ consistent and asymptotically normal. The estimator's finite sample properties are evaluated through a simulation study. 
		We also establish identification of the coefficient of the endogenous regressor in models with more general factor structures, in situations where one has access to at least two continuous measurements of the common factor.
		%  and in an empirical application, where we implement  our method to the estimation of the civic returns to college, revisiting the  work by Dee (2004).
\end{abstract}}


%\vskip 1.2in

\renewcommand{\baselinestretch}{1.2}

\normalsize
%\end{document}

%\pagebreak

\renewcommand{\thefootnote}{\arabic{footnote}} \setcounter{footnote}{0}

\setcounter{equation}{0}

\section{Introduction}

Factor models see widespread and increasing use in various areas of econometrics.
This type of structure has been employed in a variety of settings in cross sectional, panel and time series models,
and have proven to be a flexible way to model the behavior of and relationship between unobserved components of econometric models.
The basic idea behind factor models is to assume that the dependence across the unobservables is generated by a low-dimensional set of mutually independent random factors. 
The applied and theoretical research employing factor structures in econometrics is extensive. In particular, these models are often used in the treatment effect literature as a way to identify the joint distribution of potential outcomes from the marginal distributions, and then recover the distribution of treatment effects from this joint distribution.\footnote{See also \cite{AH07Handbook} for an extensive discussion of factor structures and prior studies using these models in the context of treatment effect estimation.} Factor models have been used in a number of different contexts in applied microeconomics. These include, among others, earnings dynamics \citep{AC89,BR10}, estimation of returns to schooling and work experiences \citep{ahmr17}, as well as cognitive and non-cognitive skill production technology \citep{CHS10}. \cite{hv071,hv072} provide various additional references. All of these papers, with the notable exception of \cite{CHS10}, rely on linear factor models where the unobservables are assumed to be written as the sum of a linear combination of mutually independent factors and an idiosyncratic shock. 

%Factor structures are also used in financial econometrics. In these settings it is shown that factor structures can allow for  new models for the dependence structure, or copula, of economic variables based on a latent factor structure. - see for example,  \cite{ohpattona}, \cite{ohpattonb}, \cite{hullwhite2004}. In panel data models factor models have also been useful to allow for more general forms of nonstationarity and dynamics- see, e.g. \cite{\ng}. See also \cite{stockwatson} for time series models with factor structures.

In this paper we bring together the literature on factor models with the literature on the identification and estimation of binary response models (\cite{kleinspady93,lewbel_1,ParkPhillips00,blundellpowell}), in particular triangular binary choice models (\cite{chesher3,vytlacilyildiz,shaikhvytlacil11,HV13}), by exploring the {\em informational content} of factor structures in this class of models.\footnote{See also recent work by \cite{LSZ20}, who study the identification of a triangular linear model assuming that the disturbances are related through a factor model.} Focusing on this class can be well motivated from both an empirical and theoretical perspective.
From the former, many treatment effect models fit into this framework as treatment is typically a binary and endogenous variable in the system, whose effect on outcomes is often a parameter the econometrician wishes to conduct inference on. %This is of empirical interest in many fields such as labor, industrial organization and development. 
From a theoretical perspective, inference on this type of system can be complicated, if not impossible without strong parametric assumptions, which may not be reflected in the observed data. Imposing no restriction on the structure of endogeneity often fails to achieve identification of parameter, or at  best only do so  in sparse regions of the data, thus making inference impractical in practice. In this context, modeling the endogeneity between the selection and the outcome by a factor structure may be  a useful ``in-between'' setting, which, at the very least, can be used to gauge the sensitivity of the parametric approach to their stringent assumptions.

% while desirable from a theoretical point of view because of its generality,

%Given these two extremes- the non robustness of a parametric approach and the impracticability of conducting inference with semiparametric approaches in this setting - 

We start our analysis by imposing a particular factor structure to the two unobservables in our system of binary equations described in further detail in the next section, and explore the informational content of this assumption. We assume that the unobservables from the treatment equation ($V$) and the outcome equation ($U$) are related through the following factor model: 
\begin{equation}
\label{eq1}
U=\gamma_0 V+\Pi
\end{equation}
where $\Pi$ is an unobserved random variable assumed to be distributed independently of $V$ and $\gamma_0$ is a scalar parameter. This structure generalizes the canonical case where the unobservables $(U,V)$ are jointly normally distributed, for which this relationship always holds. 
%\footnote{While this is our baseline specification, we also examine the informational content of more general factor structures involving nonparametric relationships between unobservables or multiple idiosyncratic errors.}    
Our main finding is that there is indeed informational content of factor structures in the sense that, in contrast to prior literature - notably  \cite{vytlacilyildiz} - one no longer requires an additional ``non-standard'' exclusion restriction, nor the strong support conditions on the covariates entering the outcome equation that are generally needed for identification in these models. Our identification results are constructive and translate directly into a rank based estimator of the coefficient associated with the binary endogenous variable, which we provide and study in a supplement to this paper. 

While an appealing feature of the structure considered in Equation (\ref{eq1}) is that it is a natural extension of the bivariate Probit specification that has often been considered in the literature, this model does impose significant restrictions on the nature of the dependence between the unobservables $U$ and $V$. In the paper we extend this baseline specification by considering a linear factor structure of the form: 
\begin{eqnarray}
U &=& \gamma_0 W + \eta_1\\
V &=& W + \eta_2
\end{eqnarray}
where $(W, \eta_1,\eta_2)$ are mutually independent unobserved random variables. We study the informational content of this extended factor structure in the context of triangular binary choice models and establish identification, assuming access to at least two continuous noisy measurements of the unobserved factor $W$. This setup has been used in a number of applications, in particular in labor economics. In these applications, the unobserved factor is typically interpreted as latent individual ability, about which several continuous noisy measurements are available from the data. This is the case of, for instance, \cite{CHH03}, \cite{CHS10}, \cite{HHV18} and \cite{ahmr17}, who use components of the Armed Services Vocational Aptitude Battery test as measurements of cognitive ability. %At a high level, it is interesting to note that these results complement \cite{bn10}, who show that, in the context of a linear regression model with endogenous regressors, factor models have identifying power, in that they can be used to create instrumental variables even when none of the observed variables are valid instruments. 

%Second, we provide identification results for a nonparametric factor model of the form:  
%\begin{eqnarray}
%U &=& g_0(V) + \Pi
%\end{eqnarray}
%where $g_0(.)$ is an unknown function that is assumed to satisfy some regularity conditions. 

The rest of the paper is organized as follows. In Section \ref{tmfs} we formally describe the triangular system with our factor structure, and discuss our main identification results for the parameters of interest in this model. Section \ref{sec:general} explores identification in more general factor structure models which involve multiple idiosyncratic errors, in a context where one has access to two continuous noisy measurements of the common unobserved factor. Section \ref{sec:concl} concludes. We prove Theorems \ref{thm:factorid} and \ref{thm:aux} in Sections \ref{identproof} and \ref{sec:auxpf}, respectively. In Section \ref{sec:vys}, we establish the sharp identified set of $\alpha_0$ when the support condition for point-identification is violated in the one-factor model and the necessary and sufficient condition for point-identification in the two-factor model with two continuous measurements of the common factor. The corresponding results are proved in Sections \ref{sec:sur_proof} and \ref{sec:partialid}. The Supplementary Material studies the asymptotic properties of a rank-based estimator for $\alpha_0$ and explores its finite sample properties through some Monte Carlo simulation exercises. 

Notation: throughout the paper we write $\mathbf{1}\{A\}$ to denote the usual indicator function that takes value 1 if event $A$ happens, and 0 otherwise. We also denote by $d(U)$ and $d(U|V)$ the lengths of the support of random variable $U$, and the conditional support of $U$ given $V$, respectively. 

\section{Triangular Binary Model with Factor Structure} \label{tmfs}

\subsection{Set-up and Main Identification Result} \label{setup_result}

In this section we consider the identification of the following triangular binary model:
\begin{eqnarray}\label{eq:y1-equation-binary}
Y_1 &=& {\bf 1}\{Z_1'\lambda_0+Z_3'\beta_0+\alpha_0 Y_2-U>0\}\\
Y_2 &=& {\bf 1}\{Z'\delta_0-V>0\}
\end{eqnarray}
where $Z\equiv(Z_1,Z_2)$ and $(U,V)$ is a pair of random shocks.
$Z_2$ and $Z_3$ provide the exclusion restrictions %\footnote{As we will show, an advantage of incorporating factor structures into the model is it enables
%reducing the number of exclusion restrictions. As discussed in recent work by \cite{lewbelqe}, \cite{DMZ18}, \cite{honoreselection},  it is often the case that exclusions are hard to find or to plausibly impose. An alternative approach to reducing exclusion restrictions is a partial identification approach as in \cite{honoreselection} or the assumption of nonlinearity conditions
%as in  \cite{lewbelqe}, \cite{lewbeljbes}.}
in the 
model, and the distribution of $(Z_2,Z_3)$ is required to be nondegenerate conditional on
$Z_1'\lambda_0+Z_3'\beta_0$.
We further assume that the error terms $U$ and $V$ are jointly independent of $(Z_1,Z_2,Z_3)$.
The endogeneity of $Y_2$ in (\ref{eq:y1-equation-binary}) arises when $U$ and $V$ are not independent.
% while the estimation of the model in (\ref{eq:y2-equation-binary}) is standard. 

The above model, or minor variations of it, have often been considered in the recent literature. See for example, \cite{vytlacilyildiz},
\cite{abhausmankhan}, \cite{vellaetal}, \cite{vuongxu}, \cite{khannekipelov2} and references therein. A key parameter of interest\footnote{As is always the case in models with binary outcomes, both the interpretation and the usefulness of regression coefficients warrant explanation. In the model considered here the coefficient on the treatment variable and the coefficients on exogenous variables in the binary outcome equation enable us to construct ``equivalence classes" to answer important policy questions.
	For example consider the case where  the dummy endogenous variable is job training, the exogenous regressor is years experience and the outcome variable is employment status. Knowing all coefficients would be informative on  how many additional years of experience would be needed to compensate for a lack of training so the probability of being employed stays the same.}
in our paper as is in much of the literature is $\alpha_0$. In this paper we provide conditions under which the parameters of interest are point-identified. As such, our analysis complements alternative partial-identification approaches that have been proposed in the context of triangular binary models. See, in particular, \cite{chiburis10}, \cite{shaikhvytlacil11}, and \cite{mourifie15}.\footnote{In Section \ref{sec:vys} in the supplement, we establish the sharp identified set of $\alpha_0$ when the support condition for point-identification is violated. This result highlights that, except for the fact that the sign of $\alpha_0$ is identified, we generally cannot say much about the value of $|\alpha_0|$. Related work by \cite{shaikhvytlacil11} also provides partial identification results for a triangular binary model. That the bounds for $\alpha_0$ are generally tighter in their analysis reflects the identifying power of the additional support restrictions that they impose.} As discussed in the aforementioned papers,
the parameter $\alpha_0$ is difficult, if not impossible to point identify and estimate without imposing parametric restrictions on the unobserved variables in the model, $(U,V)$.
%These parametric restrictions, such as the standard bivariate normality assumption, are not robust to misspecification in the sense that any estimator of $\alpha_0$ based on these conditions will be inconsistent if $(U,V)$ have a 
%different bivariate distribution.%\footnote{An important recent generalization of the bivariate normality assumption can be found in \cite{HV13} who propose the use of known parametric copula functions to 
%model dependence between the  two unobservables. We discuss their approach in greater detail when comparing to the factor structure approach introduce here, later in this paper. See also recent work by \cite{han_lee} who study semiparametric estimation and inference in the framework considered by \cite{HV13}.}.

The difficulty of identifying $\alpha_0$ in semi-parametric ``distribution-free" models, and the sensitivity of its identification to misspecification in parametric models is what motivates the factor structure we add in this paper to the above model. Specifically, to allow for endogeneity in the form of possible non-zero correlation between $U$ and $V$, we augment the model with the following equation:
\begin{equation}
\label{eq:factor1}
U=\gamma_0 V+\Pi
\end{equation}
where $\Pi$ is an unobserved random variable, assumed to be distributed independently of $(V,Z_1,Z_2,Z_3)$, and $\gamma_0$ is an additional unknown scalar parameter. 
%This linear, one factor structure has been imposed in the literature many times- see for example \cite{heckman_hand}.
Importantly, this type of factor structure always holds when the residuals of both equations are jointly normally distributed. % As such, this specification fits well within the ``in between'' approach (between fully non-parametric/ distribution free and parametric) that we are pursuing in this paper. 
Furthermore, this specification corresponds to the type of structure used in Independent Component Analysis (ICA), where $V$ and $\Pi$ are two mutually independent factors. This method has found many applications in various fields, including signal processing and image extraction;  applications in economics include  e.g.,  \cite{Hyvarinenetal}, \cite{Monetaetal} and \cite{Gourierouxetal}. While, in contrast to the ICA literature, the factors and the factor loadings are not the main objects of interest in our analysis, this dimension-reducing structure plays a key role in our identification results. 

Our aim is to first explore identification of the parameters
$(\alpha_0, \delta_0, \gamma_0,\beta_0,\lambda_0)$ under standard nonparametric regularity  conditions on $(V,\Pi)$. Note that the parameter $\delta_0$ in the selection equation can be identified up to scale in various ways. See, for example, \cite{kleinspady93} and \cite{MRC}, among others. We then impose the usual condition that one of $\delta_0$'s coordinates is equal to one to fix the scale. For simplicity, for the rest of the paper, we denote $X \equiv Z'\delta_0$ and assume $X$ is observed. We further define $X_1 \equiv  Z_1'\lambda_0 + Z_3'\beta_0$. However, we cannot identify $\lambda_0$ and $\beta_0$ beforehand. We propose instead to identify them along with $\alpha_0$.  
%As we explain here, neither approach generalizes the other, as the two models are non nested.

%If the marginals of $(U,V)$ are known, then our linear factor structure implies the copula between $(U,V)$ can be characterized by one parameter $\lambda$. However, comparing to \cite{HV13}, we do not require the copula to be stochastically increasing to achieve identification. If the marginals of $(U,V)$ are unknown, then

%To simplify the exposition of our strategy, in this and the following sections we will focus exclusively on the parameters $\alpha_0,\gamma_0$ and denote the linear indices by
%$X_1\equiv Z_1'\lambda_0 +Z_3'\beta_0$ and $X \equiv Z'\delta_0$, where $Z = (Z_1,Z_2)$. In particular, we treat $\delta_0$ as known. In practice, $\delta_0$ can be identified and consistently estimated in a first step using a semi-parametric single index estimator such as the one proposed by \cite{kleinspady93}.
%%This practice can be justified by established results (see, e.g. \cite{abhausmankhan},\cite{vellaetal}, \cite{khannekipelov2}) which show that these estimators are easier to identify. 
% In addition, at the end of this section, we note that we can identify $\lambda_0$ and $\beta_0$ simultaneously with $\alpha_0$. Then \eqref{eq:y1-equation-binary} and \eqref{eq:y2-equation-binary} are simplified to 
%\begin{equation}\label{eq:y1-equation-binary'}
%Y_1 = {\bf 1}\{X_1+\alpha_0 Y_2-U>0\}
%\end{equation}
%and
%\begin{equation}\label{eq:y2-equation-binary'}
%Y_2 = {\bf 1}\{X-V>0\}.
%\end{equation}

Our main identification result is based on the Assumptions {\bf A1}-{\bf A4} we state below:
\begin{description}
	\item[A1] The first coefficient of $\lambda_0$ is normalized to one so that $\lambda_0 = (1,\lambda_{0,-1}^T)^T$. The parameter $\theta_0 \equiv (\alpha_0,\gamma_0,\lambda_{0,-1},\beta_0)$ is an element of a compact subset of $\Re^{d_1+d_3+1}$, where $d_1$ and $d_3$ are the dimensions of $Z_1$ and $Z_3$, respectively. 
	\item[A2] The vector of unobserved variables, $(U,V,\Pi)$ is continuously distributed with support on a subset of $\Re^3$ and independently distributed of the vector $(Z_1,Z_2,Z_3)$.
	Furthermore, we assume that the unobserved random variables $\Pi,V$ are distributed independently of each other. 
	\item[A3] $X$ is continuously distributed with absolute continuous density w.r.t. Lebesgue measure. Its density is bounded and bounded away from zero on any compact subset of its support.
	\item[A4] Let $Z_{1,-1}$ be all the coordinates of $Z_1$ except the first one, and $d = d_1+d_3+1$.  There exist $2d$ vectors $\{z_1^{(l)},z_3^{(l)},x^{(l)}\}_{l=1}^{d}$ and $\{\tilde{z}_1^{(l)},\tilde{z}_3^{(l)},\tilde{x}^{(l)}\}_{l=1}^{d}$ in the joint support of $(Z_1,Z_3,X)$ such that   	
	\begin{align*}
	\alpha_0 + (z_{1,-1}^{(l)}-\tilde{z}_{1,-1}^{(l)})'\lambda_{0,-1} + (z_3^{(l)}-\tilde{z}_3^{(l)})'\beta_0-\gamma_0 (x^{(l)} - \tilde {x}^{(l)}) = \tilde{z}_{1,1}^{(l)}-z_{1,1}^{(l)},~l=1,\cdots,d 
	\end{align*}
	and $\text{rank}(\mathcal{M}) =d$, where 
	\begin{align*}
	\mathcal{M} = \begin{pmatrix}
	1 & \cdots & 1\\
	z_{1,-1}^{(1)} - \tilde{z}_{1,-1}^{(1)}&  \cdots & z_{1,-1}^{(d)} -\tilde{z}_{1,-1}^{(d)}\\
	z_{3}^{(1)} - \tilde{z}_{3}^{(1)}  &  \cdots & z_{3}^{(d)} - \tilde{z}_{3}^{(d)}  \\
	x^{(1)} - \tilde{x}^{(1)} &  \cdots & x^{(d)} -\tilde{x}^{(d)}\\
	\end{pmatrix}.
	\end{align*}
\end{description}

%Our main identification results for our factor structure model are based on the following conditions:
%
%\begin{description}
%\item[A1] The parameter $\theta_0\equiv(\lambda_0,\beta_0,\alpha_0,\delta_0,\gamma_0)$ is an element of a compact subset of $R^5$.
%\item[A2] The vector of unobserved variables, $(U,V,\Pi)$ is continuously distributed with support on $R^3$ and independently distributed of the vector $(Z_1,Z_2,Z_3)$.
%Furthermore, we assume the unobserved random variables $\Pi,V$ are distributed independently of each other.
%\item[A3] The vector $(Z_1,Z_2,Z_3)$ is not contained in any proper linear subspace of $R^3$ with positive probability. 
%\item[A4] The random variable $Z_2$ is continuously distributed on an interval which is a subset of $R$, conditional on all values of $\tilde Z=Z_1'\lambda_0+Z_3'\beta_0$.
%\item[A5]   $|\alpha_0| < \ell(Z_1'\lambda_0+Z_3'\beta_0 - \gamma_0 Z'\delta_0)$ where $\ell(\cdot)$ denotes the length operator.
%\end{description}

Before turning to our main identification result, a couple of remarks are in order. %But what best demonstrates the identifying power of the factor structure is the comparison of our other assumptions  to those typically imposed in the literature for this model without the factor structure.
%As explained in the remarks below, notably Remark \ref{crucialremark},  the factor structure enables the relaxation of strong exclusion and support conditions typically assumed for inference in these types of models.

\begin{remark}
	The first part of Assumption A1 is a standard scale normalization. Assumption A2 is also standard in this literature. % in both the unobservables $U,V$ as well as the independence between $\Pi$ and $V$.        
	The assumption that the instruments are independent of the unobservables can also be found in, among others, \cite{abhausmankhan}, \cite{vytlacilyildiz}, \cite{vellaetal}, and \cite{khannekipelov2}. The assumption of independence between $\Pi$ and $V$ is also made in \cite{baing} and \cite{CHH03}.
\end{remark}

\begin{remark}
	Assumptions A3 and A4 impose some restrictions on the distributions of the covariates entering the selection and outcome equations, respectively. Specifically, Assumption A3 requires one component of the covariates $Z$ entering the selection equation to be continuously distributed, which is often required in models with discrete outcomes. In contrast, Assumption A4 only requires some variation of $(Z_1,Z_3)$. In particular, the distribution of $(Z_1,Z_3)$ cannot be degenerate but is allowed to be discrete. This assumption can be interpreted as a full rank condition, which ensures that the system of linear equations that delivers point identification has a unique solution.
	%Recent papers - e.g. \cite{dhault2014},
	%\cite{torgo2014} establish identification with discrete instruments, but in a different setup where the endogenous %variable in the outcome equation is continuously distributed.
	\label{rem:A34}
\end{remark}

We now turn to our main identification result, Theorem \ref{thm:factorid}, which concludes that under our stated conditions and our factor structure we can attain point identification of the vector of parameters $\theta_0$. 
%Under this set of conditions, we have the following identification result.
% \textbf{Yichong: We did not prove the following theorem in the appendix. In fact, we only consider a simplified case in which $(\lambda_0,\beta_0,\delta_0)$ are known.}

\begin{theorem}
	Under Assumptions A1-A4, $\theta_0$ is point identified.
	\label{thm:factorid}
\end{theorem}

An important takeaway from this result, which we discuss further in Subsection~\ref{subsec:lit} below, is that imposing the factor structure~($\ref{eq:factor1}$) yields point-identification under weaker support conditions when compared to the existing literature, and does not require a second exclusion restriction either. In particular, our model delivers point-identification of the parameters of interest even in situations where all of the regressors from the outcome equation are discrete. This indicates that, from the selection equation combined with the factor structure that we impose here, we can overturn the non-identification result of \cite{bierenshartog} which would apply to the outcome equation alone.%\footnote{See \cite{tamer_manski} and \cite{magnac-maurin08} who provide set-identification results for semi-parametric regression models in the presence of discrete or interval-valued regressors. }

%\begin{rem}
The proof of Theorem \ref{thm:factorid}, which is reported in Section \ref{identproof} in the Supplementary Appendix, relies on the fact that, for two observations $(Z_1,Z_3,X)$ and $(\tilde{Z}_1, \tilde{Z}_3, \tilde{X})$, 
\begin{align}
\label{eq:match}
& \partial_x P^{11}(Z_1,Z_3,X)/f_V(X) +\partial_x P^{10}(\tilde {Z}_1, \tilde{Z}_3,\tilde X)/f_V(\tilde X) =0  \notag \\
\iff &\alpha_0 + (Z_1-\tilde{Z}_1)'\lambda_0 + (Z_3-\tilde{Z}_3)'\beta_0-\gamma_0 (X - \tilde X)=0,
\end{align} 
where $f_V(\cdot)$ is the pdf. of $V$, which is identified over the support of $X$, and $P^{ij}(z_1,z_3,x) \equiv Prob(Y_1 = i,Y_2 = j|Z_1 =z_1,Z_3 = z_3,X=x)$ ($\partial_x P^{ij}(z_1,z_3,x) $) denote the choice probability (partial derivative of the $ij$-choice probability with respect to the third argument), which are both identified from the data. 

\begin{rem} 
	This identification result can be extended to the case of a separable nonparametric factor model. Namely, consider the following relationship between unobserved components:
	\begin{equation}
	U=g_0(V)+\tilde \Pi
	\end{equation}
	where $\tilde \Pi$ is an unobserved random variable assumed to be distributed independently of $V$ and all instruments.
	$g_0(\cdot)$ is an unknown function assumed to satisfy standard smoothness conditions.
	The parameter of interest is $(\alpha_0,\lambda_0,\beta_0)$, but now the unknown nuisance parameter in the factor equation is infinite dimensional. By replacing $\gamma_0 X$ by $g_0(X)$ in \eqref{eq:match}, we have 
	\begin{align}
	\label{eq:matchg}
	& \partial_x P^{11}(Z_1,Z_3,X)/f_V(X) +\partial_x P^{10}(\tilde {Z}_1, \tilde{Z}_3,\tilde X)/f_V(\tilde X) =0  \notag \\
	\iff &\alpha_0 + (Z_1-\tilde{Z}_1)'\lambda_0 + (Z_3-\tilde{Z}_3)'\beta_0-(g_0(X) - g_0(\tilde X))=0. 
	\end{align} 
	One can then establish identification after modifying the rank condition A4 by replacing $\gamma_0 (x^{(l)} - \tilde {x}^{(l)})$ by $g_0(x^{(l)})-g_0(\tilde {x}^{(l)})$.
	%\textbf{Comments: do we need to mention the estimation here. If we do, then, we also need to mention the estimation of the one-factor model in the previous section.}
	
	%For estimation, now our approach is to replace the vector $X$ with a series of basis functions of $X$, such as, for example orthonormal polynomials, in $X$.
	%Those basis functions are meant to serve as an approximation of $g_0(\cdot)$. With that replacement, we carry exactly as before, except now instead of estimating a kernel weighted linear regression model we estimate a kernel weighted semi linear regression model as in, e.g., \cite{robinson}. Section \ref{semilinear} in the Supplementary Appendix  provides details of how to construct such an estimator and outlines its large sample properties.
	
\end{rem}

%\begin{rem}
%\cite{vytlacilyildiz} show in their Appendix B that a sufficient condition for their identification result without the factor structure is 
%\begin{align}
%\label{eq:vy}
%\Supp(Z_3'\beta_0 + \alpha_0) \cap \Supp(Z_3'\beta_0 ) \neq \emptyset
%\end{align}
%In Appendix \ref{sec:vys}, we show that, in a ``condensed" setting, such type of condition is also necessary. In Appendix \ref{sec:vys}, we further show the extra region in the parameter space that has identification given the fact structure and doesn't without. 
%\end{rem}

%\begin{rem}
%Note that $\alpha_0$ is over-identified with or without the factor structure, provided that the corresponding support conditions hold. Both \cite{vytlacilyildiz} and this paper propose $\sqrt{n}$-consistent estimators of $\alpha_0$.\footnote{See Section \ref{asymprop} in the supplement for more details.} The key benefit of imposing the factor structure is not to improve the convergence rate of the estimator, but rather to relax the support condition needed for point-identification of $\alpha_0$.  
%\end{rem}

%\begin{rem}
%\cite{shaikhvytlacil} and \cite{mourifie15} discuss the partial identification of the binary triangular system without the factor structure. Although our paper mainly focuses on point identification, in the supplement, we
%\end{rem}

\begin{rem}
	\label{rem:rank_similarity}
	We assume rank invariance in \eqref{eq:factor1}. It is possible to relax such condition to rank similarity.\footnote{We thank the referee for pointing this out.} Specifically, we can consider the following model: 
	\begin{eqnarray*}\label{eq:y1-equation-binary'}
		Y_1 &=& {\bf 1}\{Z_1'\lambda_0+Z_3'\beta_0+\alpha_0 Y_2-U(Y_2)>0\}\\
		Y_2 &=& {\bf 1}\{Z'\delta_0-V>0\},
	\end{eqnarray*}
	where
	\begin{align*}
	U(y_2) = \gamma_0 V + \Pi(y_2), ~\text{for}~y_2 = 0,1.
	\end{align*}
	We further assume $(V,\Pi(1),\Pi(0))$ is continuously distributed with support on a subset of $\Re^3$ and independently distributed of the vector $Z_1,Z_2,Z_3$, $V$ and $(\Pi(1),\Pi(0))$ are independent, $P(\Pi(1) \leq \pi) = P(\Pi(0) \leq \pi)$ for $\pi \in \Re$, and Assumptions {\bf A1},  {\bf A3}, and  {\bf A4} hold. Then, we can identify $\theta_0$ by a similar argument as the proof of Theorem \ref{thm:factorid}. 
\end{rem}

\subsection{Connection with Prior Literature} \label{subsec:lit}

We now discuss in detail how our setup and main identification result relates to the existing literature. 

In a related work, \cite{HV13} consider the identification of a generalized bivariate Probit model.\footnote{See also recent work by \cite{han_lee} who study semiparametric estimation and inference in the framework considered by \cite{HV13}.} Our linear factor structure and the one-parameter copula model considered in \cite{HV13} are not nested by each other. First, note that based on the factor structure, we can recover $F_\Pi$, the distribution of $\Pi$, as a function of $(F_U,F_V,\gamma_0)$ by deconvolution. We can then write the copula of $(U,V)$ as 
$$F_{U,V}(F_U^{-1}(u),F_V^{-1}(v)) = \int_{-\infty}^{F_V^{-1}(v)}F_\Pi(F_U^{-1}(u) - \gamma_0 w; F_U,F_V,\gamma_0)f_V(w)dw = C(u,v;F_U,F_V,\gamma_0).$$  
The copula depends not only on $\gamma_0$ but also on two infinite dimensional parameters $(F_U,F_V)$. Thus, unlike \cite{HV13}, our factor structure cannot be characterized by a one-parameter copula. In addition, in order to achieve identification, \cite{HV13} first nonparametrically identify the two marginals by assuming the existence of a full support regressor that is common to both equations.\footnote{\cite{HV13} establish their identification of the coefficient on the endogeneous regressor (Theorems 4.2 and 5.1) under the assumption that the marginal distributions $F_\eps$ and $F_\nu$ are known. Then, they verify this condition by showing the identification of these two marginal distributions using large support common regressors.} In contrast, our approach does not rely on the existence of such a regressor. Under the factor structure assumed in our analysis, we bypass the nonparametric identification of the marginals as a whole and directly consider the identification of the structural parameters. It follows that our model cannot be nested by the one-parameter copula model considered by \cite{HV13}. On the other hand, there exist one-parameter copula models that cannot be decomposed into linear factor structures.\footnote{For instance, suppose that $(U,V)$ has a Gaussian copula with correlation $\rho$, and that the marginal distributions of $U$ and $V$ are uniform $[0,1]$. It then follows that, denoting by $\Phi(.)$ the standard normal cdf., $\left(\Phi^{-1}(U), \Phi^{-1}(V)\right)$ is bivariate normal with correlation $\rho$, which in turn yields the following non-linear relationship between $U$ and $V$: $U = \Phi\left( \rho \Phi^{-1}(V) + W\right)$, where $W$ is normally distributed and independent from $V$.} This implies that our model does not nest \cite{HV13} either.

%Our approach, which consists in adding more structure to the fully semiparametric triangular binary system and quantify the identifying power of the added structure is, in one sense, the reverse approach of generalizing the fully parametric model. Such an approach has been taken recently in \cite{HV13}, who begin with a bivariate Probit model, and generalize it with the introduction of a class of one parameter copulas, providing conditions such that identification can still be obtained.

Our analysis also relates to \cite{vytlacilyildiz} and \cite{vuongxu}, who consider the identification of $\alpha_0$ in a triangular binary model. Our identification result, however, differs from theirs in important ways. Namely, denote $X = Z'\delta_0 = Z_1'\delta_{1,0} + Z_2'\delta_{2,0}$. Then, Assumption A4 implies that we can find a pair of observations $(z_1,z_2,z_3)$ and $(\tilde{z}_1,\tilde{z}_2,\tilde{z}_3)$ such that 
\begin{align}
\label{eq:I1}
z_1'\lambda_0 + z_3'\beta_0 +\alpha_0 - \gamma_0(z_1'\delta_{1,0} + z_2'\delta_{2,0}) = \tilde{z}_1'\lambda_0 + \tilde{z}_3'\beta_0  - \gamma_0(\tilde{z}_1'\delta_{1,0} + \tilde{z}_2'\delta_{2,0}).
\end{align} 
In contrast, using our notation, \cite{vytlacilyildiz} require that one can find a pair of observations $(z_1,z_2,z_3)$ and $(\tilde{z}_1,\tilde{z}_2,\tilde{z}_3)$ such that $z'\delta_{0} = \tilde{z}'\delta_0$ and  
\begin{align}
\label{eq:I2}
z_1'\lambda_0 + z_3'\beta_0 +\alpha_0  = \tilde{z}_1'\lambda_0 + \tilde{z}_3'\beta_0.
\end{align} 
\cite{vuongxu} do not assume the existence of $Z_3$. In our binary outcome setup, the functions $h(0,x,\tau)$ and $h(1,x,\tau)$ defined in \cite{vuongxu} are equal to $1\{x + F_{-U}^{-1}(\tau) \geq 0\}$ and $1\{x + \alpha+F_{-U}^{-1}(\tau) \geq 0\}$, respectively, where $x = z_1'\lambda_0$ and $F_{-U}$ is the CDF of $-U$. Then, \citet[Assumption C'(ii)]{vuongxu} requires that we can find $z_1$ and $\tilde{z}_1$ in the support of $Z_1$ so that for any $\tau_1,\tau_2$, if $1\{\tilde{z}_1'\lambda_0 + F_{-U}^{-1}(\tau_1) \geq 0\} = 1\{\tilde{z}_1\lambda_0 + F_{-U}^{-1}(\tau_2) \geq 0\}$, then $1\{z_1'\lambda_0 + \alpha_0 + F_{-U}^{-1}(\tau_1) \geq 0\} = 1\{z_1\lambda_0 + \alpha_0 + F_{-U}^{-1}(\tau_2) \geq 0\}$. Provided that the support of $U$ nests the supports of $Z_1'\lambda_0$ and $Z_1'\lambda_0+\alpha_0$, \citet[Assumption C'(ii)]{vuongxu} is then equivalent to:\footnote{To see this, note that if, say, $z_1'\lambda_0 +\alpha_0  > \tilde{z}_1'\lambda_0$, then we can find $\tau_1,\tau_2$ such that $-z_1'\lambda_0 - \alpha_0 \leq F_{-U}^{-1}(\tau_1) < -\tilde{z}_1'\lambda_0$ and $F_{-U}^{-1}(\tau_2) < -z_1'\lambda - \alpha_0 < -\tilde{z}_1'\lambda_0$. This violates the above requirement, and thus, shows that \citet[Assumption C'(ii)]{vuongxu} implies \eqref{eq:I3}. On the other hand, if $z_1'\lambda_0 +\alpha_0  = \tilde{z}_1'\lambda_0$, then \citet[Assumption C'(ii)]{vuongxu} holds trivially. } 
\begin{align}
\label{eq:I3}
z_1'\lambda_0 +\alpha_0  = \tilde{z}_1'\lambda_0.
\end{align}

Several remarks are in order. First, note that sufficient support conditions for the restrictions \eqref{eq:I1}--\eqref{eq:I3} are $d(Z_1'\lambda_0 + Z_3'\beta_0 - Z'\delta_{0}\gamma_0) \geq |\alpha_0|$, $d(Z_1'\lambda_0 + Z_3'\beta_0|Z'\delta_0)\geq |\alpha_0|$, and $d(Z_1'\lambda_0 |Z'\delta_0)\geq |\alpha_0|$ with a positive probability, respectively, where $d(\cdot)$ denotes the ``length" of its argument. These three support conditions are such that $$d(Z_1'\lambda_0 + Z_3'\beta_0 - Z'\delta_{0}\gamma_0) \geq d(Z_1'\lambda_0 + Z_3'\beta_0|Z'\delta_0) \geq d(Z_1'\lambda_0|Z'\delta_0),$$ where the first and second inequalities are strict if $Z_2$ and $Z_3$ have at least one continuous component, respectively. Importantly, we show in Section \ref{sec:vys} of the Supplement that for a version of the triangular binary model with univariate $Z_2$ and $Z_3$ and no common regressor $Z_1$, the support condition $d(Z_1'\lambda_0 + Z_3'\beta_0|Z'\delta_0)\geq |\alpha_0|$ is 
actually also necessary to the identification of the model without factor structure. This implies that by imposing our factor structure, one can identify values of $\alpha_0$ in a region that cannot be identified in the model considered by \cite{vytlacilyildiz}. Such region is characterized in Section \ref{sec:vys} of the Supplement. 

Second, it directly follows from these support conditions that, in the presence of a factor model and in contrast to both \cite{vytlacilyildiz} and \cite{vuongxu}, variation in $Z_2$ helps in the identification of $\alpha_0$. In that sense, the factor model allows to restore the intuition from standard IV approaches in linear models that variation in the instrument $Z_2$ is critical to the identification of the parameters of the outcome equation. Related to this, the support of $Z_2$ plays an important role in our identification analysis. In particular, if $Z_2$ is discrete, our identification strategy requires sufficient variation in the variables in the outcome equation, namely $Z_1$ and $Z_3$. In this case, our support requirement is equivalent to that assumed by \cite{vytlacilyildiz}. %As long as their identification condition holds, one also can achieve identification. 

Third, another important aspect of Assumption A4 is that it does not impose any constraint on the variables from the outcome equation. Specifically, consider a case where the outcome equation does not contain a variable that is excluded from the selection equation (i.e., $\beta_0=0$), the regressor that is common to both equations, $Z_1$, is scalar and binary, and where $\lambda_0 = 1$. In this case, one can show that the identifying support conditions associated with \cite{vytlacilyildiz} \eqref{eq:I2} and \cite{vuongxu} \eqref{eq:I3} generally fail to hold, except for a finite set of values $\alpha_0  \in \{-1,0,1\}$. In contrast, our support restriction \eqref{eq:I1} holds under more general conditions: without any restriction on $\alpha_0$ if one element of $Z_2$ is continuous with large support, and on a continuum of possible values for $\alpha_0$ if one element of $Z_2$ is continuous with bounded support. In that sense, the factor structure replaces the need for a continuous component in $(Z_1,Z_3)$ in the outcome equation. %, distinguishing our identification result from those in \cite{vytlacilyildiz} and \cite{vuongxu}.

Finally, at a high level, our identification strategy shares similarities with the Local Instrumental Variable (LIV) approach that has been proposed by \cite{hv05} and further discussed by \cite{cl09}. In particular, our identifying restriction~\eqref{eq:match} can be alternatively derived from a local IV strategy applied to a potential outcomes model characterized by $Y_1(y_2)={\bf 1}\{Z_1'\lambda_0+Z_3'\beta_0+\alpha_0 y_2-U>0\}$, with treatment given by $Y_2= {\bf 1}\{Z'\delta_0-V>0\}$. In contrast to the LIV literature though, we focus in our analysis on the structural parameter $\alpha_0$ rather than on the marginal treatment effects. Our identification result shows that, by leveraging the identifying power of the factor structure, one can identify $\alpha_0$ under weaker support restrictions than in the prior literature. In particular, our strategy makes it possible to use variation in $X = Z'\delta_0$ to identify $\alpha_0$, even when all the components of $Z_1$ and $Z_3$ are discrete.\footnote{An alternative approach to identifying this parameter can be found in \cite{lewbel_1}. In his approach a second equation to model the endogenous variable is not needed, nor is the factor structure we impose. However, he imposes a strong support condition on a variable like $Z_3$ requiring that it exceeds the length of the unobservable $U$.} 

%As explained in \cite{khantamer}, such an approach precludes even bounding $\alpha_0$ if the support condition on $Z_3$ is not satisfied.}  

\section{Extended Factor Structure in the presence of Continuous Measurements} \label{sec:general}\setcounter{equation}{0}
Up until now we have proposed identification and estimation results for a triangular system with a particular factor structure. A disadvantage of this structure is that it only includes one idiosyncratic shock ($\Pi$). We consider below an extension that addresses this limitation. 
%The other is the linear in parameter relationship between the two unobserved components, which leaves open the possibility of misspecification. 

%The specific conditions and asymptotic distribution theory is still to be completed.

%\subsection{Identification with Two Idiosyncratic Shocks and Two Continuous Measurements}

Namely, we consider the following model: 
\begin{equation}
\begin{aligned}
& Y_1 = \mathbf{1}\{X_1 + \alpha_0 Y_2 - U \geq 0 \} \\
& Y_2 = \mathbf{1}\{X - V \geq 0 \},
\end{aligned}
\label{eq:2Fdis}
\end{equation}
where $X_1 = Z_1'\lambda_0 + Z_3'\beta_0$, $X =  Z'\delta_0$, $U = \gamma_0 W + \eta_1$, $V = W + \eta_2$, and $(W, \eta_1,\eta_2)$ are mutually independent. In this setup, $W$ can be interpreted as an unobserved confounder that satisfies the matching-on-unobservables condition $(Y_1(0),Y_1(1)) \indep Y_2 | W,X,X_1$ \citep{AH07Handbook}. Recall that, following the arguments in Section~\ref{setup_result} above, we assume that $X$ is observed. In addition, we assume two auxiliary continuous measurements 
\begin{align}
& Y_3 = \nu_0 W + \eta_3 \notag \\
& Y_4 = \sigma_0 W + \eta_4, 
\label{eq:measurements}
\end{align}
where $(W, \eta_1,\eta_2,\eta_3,\eta_4)$ are mutually independent, and $\nu_0 \neq 0.$\footnote{In practice, the continuous measurements might also depend on some observable characteristics. Our analysis goes through in this case after residualizing $Y_3$ and $Y_4$.}

Our identification result is based on the following assumptions: 

\begin{description}
	\item[B0] The first coefficient of $\lambda_0$ is normalized to one so that $\lambda_0 = (1,\lambda_{0,-1}^T)^T$. The parameter $\theta_0 \equiv (\alpha_0,\gamma_0,\lambda_{0,-1},\beta_0,\nu_0,\sigma_0)$ is an element of a compact subset of $\Re^{d_1+d_3+3}$, where $d_1$ and $d_3$ are the dimensions of $Z_1$ and $Z_3$, respectively. The vector of unobservables in the outcome and selection equations $(W,\eta_1,\eta_2,\eta_3)$ are independently distributed of the vector $(Z_1,Z_2,Z_3)$.  Both $\eta_1$ and $\eta_2$ are continuously distributed. 
	\item[B1] $\gamma_0 \neq 0$. $X$ is continuously distributed with absolute continuous density w.r.t. Lebesgue measure over the whole real line, conditionally on $Z_1$ and $Z_3$. The unconditional density of $X$ is bounded and bounded away from zero on any compact subset of its support. 
	\item[B2] $W$ is not normally distributed or both $\eta_3$ and $\eta_4$ do not have a Gaussian component. 
	\item [B3] $E(\eta_3) = E(\eta_4) = 0$, $E(|\eta_3|)< \infty$, and $E(|\eta_4|)< \infty$. 
	\item [B4]  $E(\exp(i \zeta \eta_2))$, $E(\exp(i \zeta \eta_3))$, and $E(\exp(i \zeta \eta_4))$ do not vanish for any $\zeta \in \Re$, where $i = \sqrt{-1}$. 
	\item [B5]  $E(\exp(i \zeta W)) \neq 0$  for all $\zeta$ in a dense subset of $\Re$. 
	\item [B6] The distributions of $W$, $\eta_2$, and $\eta_3$ admit uniformly bounded densities $f_W(\cdot)$, $f_{\eta_2}(\cdot)$, and $f_{\eta_3}(\cdot)$ with respect to the Lebesgue measure that are supported on an interval (which may be infinite), respectively.
	\item [B7] Let $Z_{1,-1}$ be all the coordinates of $Z_1$ except the first one, and $d = d_1+d_3+1$. There exist $2d$ vectors $\{z_1^{(l)},z_3^{(l)}\}_{l=1}^{d}$ and $\{\tilde{z}_1^{(l)},\tilde{z}_3^{(l)}\}_{l=1}^{d}$ in the joint support of $(Z_1,Z_3)$ and $\{w^{(l)}\}_{l=1}^d,\{\tilde{w}^{(l)}\}_{l=1}^d$ such that   	
	\begin{align*}
	\alpha_0 + (z_{1,-1}^{(l)}-\tilde{z}_{1,-1}^{(l)})'\lambda_{0,-1} + (z_3^{(l)}-\tilde{z}_3^{(l)})'\beta_0-\gamma_0 (w^{(l)} - \tilde {w}^{(l)}) = \tilde{z}_{1,1}^{(l)} - z_{1,1}^{(l)},~l=1,\cdots,d 
	\end{align*}
	and $\text{rank}(\mathcal{M}) =d$, where 
	\begin{align*}
	\mathcal{M} = \begin{pmatrix}
	1 & \cdots & 1\\
	z_{1,-1}^{(1)} - \tilde{z}_{1,-1}^{(1)}&  \cdots & z_{1,-1}^{(d)} -\tilde{z}_{1,-1}^{(d)}\\
	z_{3}^{(1)} - \tilde{z}_{3}^{(1)}  &  \cdots & z_{3}^{(d)} - \tilde{z}_{3}^{(d)}  \\
	w^{(1)} - \tilde{w}^{(1)} &  \cdots & w^{(d)} -\tilde{w}^{(d)}\\
	\end{pmatrix}.
	\end{align*}
\end{description}
We now discuss these assumptions, before turning to the identification result. First, Assumption B0 is similar to Assumptions A1 and A2. We only need one of the idiosyncratic errors in the continuous measurements to be independent of the covariates because the other one is used to identify the distribution of the common factor $W$ only. Second, as we assume in Assumption B1 that $\gamma_0 \neq 0$ and $X$ has full support, the support condition 
$$d(Z_1'\lambda_0  + Z_3'\beta_0 - \gamma_0 X) \geq |\alpha_0|.$$
holds automatically. The full support condition of $X$ is necessary to identify the density of $V$, which is further used to identify the distribution of $\eta_2$. Assumption B1 reinforces this condition by supposing that $X$ has full support conditional on $Z_1$ and $Z_3$, which is needed to identify the parameters from the outcome equation in a second step. Since $X =  Z'\delta_0$ with $Z=(Z_1,Z_2)$, this is in turn equivalent to $Z_2$ having full support conditional on $Z_1$ and $Z_3$. Third, Assumptions B2--B6 imply Assumptions 1 to 4 in \cite{SH13}. In practice we add the condition that the characteristic function of $\eta_2$ does not vanish, which is used for the deconvolution arguments in the proof of Theorem \ref{thm:aux}. We refer the reader to \cite{SH13} for more discussions of these assumptions.\footnote{Note that \citet[Assumptions 5 and 6]{SH13} hold automatically in our model with $\nu_0 \neq 0.$}

\begin{theorem}
	\label{thm:aux}
	If \eqref{eq:2Fdis}--\eqref{eq:measurements} and Assumptions {\bf B0}--{\bf B7} hold, then $\theta_0$ are identified. 
\end{theorem}

The proof of Theorem \ref{thm:aux} can be found in Section \ref{sec:auxpf} of the Supplement. Several remarks are in order. First, while we allow for a more general factor structure on the unobservables $U$ and $V$, we also depart from our baseline specification by supposing that we have access to two continuous noisy measurements of the common factor $W$. This is a standard requirement in the nonparametric measurement error literature \citep{hs08}. Besides, assuming access to a set of (selection-free) noisy measurements of the unobserved factors is also very standard in the evaluation literature. See, among many others, \cite{CHH03}, \cite{HN07}, \cite{hv071}, and \cite{CHS10}. 

For instance, in applications in labor economics, the unobserved factor $W$ often captures individual ability. This would apply, for example, to the evaluation of the effect of employment while in college ($Y_2$) on college graduation ($Y_1$). In this example, natural candidates for $Z_2$ are local labor market variables, including average wages and unemployment rate, while candidates for $Z_1$ include, among others, eligibility to financial aid programs providing tuition subsidy to students who maintain a minimum level of academic achievement.\footnote{See \cite{ScottClaytonJHR11} for an evaluation of a program of this kind (PROMISE scholarship in West Virginia), and for a discussion of similar merit-based scholarship programs in place in other states.} In this context, cognitive skill measurements, such as the ASVAB test components that are available in the NLSY79 and NLSY97 surveys, are natural and often used candidates for the continuous measurements $(Y_3,Y_4)$ \citep{ahmr17}.

%while candidates for the regressors $Z_1$ that are common to both equations include field of study and family background variables such as family income and family size (see, e.g., \citeauthor{BFM13}, \citeyear{BFM13}). 

%Second, one can view $Y_1$ and $Y_2$ as two measurements of the $W$ with errors $\eta_1$ and $\eta_2$. \cite{hs08} shows in order to identify the distribution of $W$, at least three measurements are needed. In addition, they require two among the three measurements must be continuous. In our case, since both $Y_1$ and $Y_2$ are discrete, we at least need two more continuous measurements $Y_3$ and $Y_4$. We are able to show identification in this case. 

Second, as is clear from the proof of Theorem \ref{thm:aux}, the key purpose of the continuous measurements is to identify the distribution of the common factor $W$. While we assume in this section that the measurement equations are linear, it is possible to identify $\theta_0$ with a more general nonlinear system of continuous measurements, provided that the researcher has access to at least three such measurements. One can then combine Theorem 2 in \cite{CHS10} (Section 3.3, pp. 894-895), that yields identification of the distribution of $W$, with the proof of Theorem~\ref{thm:aux} in order to show identification of $\theta_0$ for the case of nonlinear auxiliary measurements. Assuming access to a set of at least three measurements also makes it possible to relax the non-normality requirement imposed in Assumption B2.

Third, under the previous set of assumptions, the average treatment effect (ATE) is also identified. Key to this identification result is the full support condition on $X$ given $Z_1$ and $Z_3$ (Assumption B1). Note that the conditional ATE given $X_1=x_1$ is equal to $F_U(x_1 + \alpha_0) - F_U(x_1)$. In addition, 
$$P(Y_1 = 1,Y_2 =1|X_1=x_1, X=x) = F_{U,V}(x_1 + \alpha_0, x).$$
One can let $x \rightarrow \infty$ so that   
$$\lim_{x \rightarrow \infty}P(Y_1 = 1,Y_2 =1|X_1=x_1, X=x) = F_U(x_1+\alpha_0).$$
Similarly, 
$$\lim_{x \rightarrow -\infty}P(Y_1 = 1,Y_2 =0|X_1=x_1, X=x) = F_U(x_1).$$
This identifies the conditional and unconditional ATE. 

Fourth, similar to the earlier discussions in Remark \ref{rem:A34} and Section \ref{subsec:lit}, Assumption B7 may still hold even when $Z_3$ is an empty set and $Z_1$ is discrete, since $W$ is assumed to have full support. In such a case, identification primarily relies on the factor structure and the variation of the covariates in the selection equation, rather than that in the outcome equation. In this respect, this identification result is similar in spirit to Theorem \ref{thm:factorid} and different from the existing identification results in the literature for triangular binary models, e.g., \cite{vytlacilyildiz} and \cite{vuongxu}. More generally, in Section \ref{sec:indep2} in the supplement we establish that the factor model provides identification restrictions that are not otherwise available.\footnote{Specifically, we consider a version of the model \eqref{eq:2Fdis}, where we do not impose the factor structure and allow for an arbitrary (unknown to econometricians) dependence structure across the unobservables of the model. In this case, we show non-identification of $\alpha_0$ as long as $|\alpha_0| > b-a$, where $[a,b]$ denotes the conditional support of $X_1$ given $X$ and, consistent with our Assumption B1, $X$ has full support on the real line. However, by imposing the factor structure (and other conditions implied by B0--B7), Theorem \ref{thm:aux} shows that $\alpha_0$ is identified for this model even when $|\alpha_0|>b-a$.}

%Finally, in Section \ref{sec:indep2} in the supplement, we consider the model \eqref{eq:2Fdis} with two measurements $Y_3$ and $Y_4$. We assume $X_1$ and $X$ are observed and the conditional support of $X_1$ given $X$ is an interval denoted as $[a,b]$. The support of $X$ is allowed to be the whole real line. We do not impose the factor structure and allow for an arbitrary (unknown to econometricians) dependence structure among $(U,V,Y_3,Y_4)$. In this case, we show the non-identification of $\alpha_0$ as long as $|\alpha_0| > b-a$. However, by imposing the factor structure (and other conditions in B0--B7), Theorem \ref{thm:aux} shows $\alpha_0$ is identified for this model even when $|\alpha_0|>b-a$. This shows the two-factor model provides identification restrictions that are not otherwise available.  

Finally, we can relax the rank invariance condition to rank similarity by replacing $Y_1 = 1\{X_1 + \alpha_0 Y_2 - U \geq 0\}$ by $Y_1 = 1\{X_1 + \alpha_0 Y_2 - U(Y_2) \geq 0\}$. We then require $U(y_2) = \gamma_0 W + \eta_1(y_2)$ for $y_2 = 0,1$. If Assumptions {\bf B0}--{\bf B7} hold with $\eta_1$ replaced by $(\eta_1(1),\eta_1(0))$ and $P(\eta_1(1) \leq e) = P(\eta_1(0) \leq e)$ for $e \in \Re$, then we can still identify $\theta_0$ by a similar argument as the proof of Theorem \ref{thm:aux}.

%\subsection{Nonparametric  Factor Model}
%We leave the exploration to multi factor models to future research and focus in this section on a single, but nonlinear, nonparametric factor structure. As we can show here, the approach taken in the previous sections can readily extend to the more general model.

\section{Conclusion}
\label{sec:concl}
In this paper, we explore the identifying power of linear factor structures in the context of simultaneous binary response models. We impose two alternative types of factor structures on the unobservables of the model. The first setup is a natural distribution-free extension of the bivariate Probit model, while the second model corresponds to a standard linear factor model with one common factor and two equation-specific idiosyncratic shocks. We establish that both factor models have identifying power in that they make it possible to relax some of the exclusion and support conditions typically required for identification in this class of models (\citeauthor{vytlacilyildiz}, \citeyear{vytlacilyildiz}). Overall, our analysis adds to our understanding of the identifying power of factor models, beyond their well known usefulness to recover the joint distribution of potential outcomes from the marginal distributions. 

%complements results obtained by \cite{bn10} in the context of a linear regression model with endogenous regressors, and, more generally, 

%We found that for a binary-binary system the factor structure we considered did indeed add informational content. Specifically, it enabled the relaxation of both the exclusion and support conditions typically employed in the identification of these models. %As we then demonstrated factor structures then enabled the regular identification of parameters of interest, and we proposed a new rank based estimation procedure that converged at a parametric rate with a limiting normal distribution. Finite sample properties of the estimator were demonstrated through simulation studies.

The work here opens areas for future research. The factor structure we assume could prove useful in more general nonlinear models. For instance, non-triangular discrete systems have shown to be an effective way to model entry games in the empirical industrial organization literature- see, for example, \cite{tamer_2}. However, as shown in \cite{khannekipelov2}, identification of structural parameters in these models can be even more challenging than for the triangular model
considered in this paper, and furthermore, as shown recently in \cite{khannekipelov3}, conducting valid uniform interest in all these models is very difficult. It would be useful to determine if factor structures on the unobservables could alleviate this problem. We leave this open question to future work.
% \newpage

\appendix
\section{Proof of Theorem \ref{thm:factorid}}\label{identproof}
\setcounter{equation}{0}
{\bf Proof:} Note that
\begin{equation*}
\begin{aligned}
P^{11}(z_1,z_3,x) & = \int_{-\infty}^{x}F_\Pi(z_1'\lambda_0 + z_3'\beta_0 + \alpha_0 - \gamma_0 v)f_V(v)dv \\
P^{10}(\tilde{z}_1,\tilde{z}_3,\tilde{x}) & = \int^{+\infty}_{\tilde{x}}F_\Pi(\tilde{z}_1'\lambda_0 + \tilde{z}_3'\beta_0 - \gamma_0 v)f_V(v)dv.
\end{aligned}
\end{equation*}
Taking derivatives w.r.t. the third argument of the LHS function, we obtain
\begin{equation*}
\begin{aligned}
\partial_x P^{11}(z_1,z_3,x)/f_V(x) & = F_\Pi(z_1'\lambda_0 + z_3'\beta_0 + \alpha_0 - \gamma_0 x) \\
- \partial_x P^{10}(\tilde{z}_1,\tilde{z}_3,\tilde{x})/f_V(\tilde{x}) & = F_\Pi(\tilde{z}_1'\lambda_0 + \tilde{z}_3'\beta_0 - \gamma_0 \tilde{x}).
\end{aligned}
\end{equation*}	
By Assumption \textbf{A4}, we know that there exists pairs such that	
$$Z_1'\lambda_0 + Z_3'\beta_0 + \alpha_0 - \gamma_0 X = \tilde{Z}_1'\lambda_0 + \tilde{Z}_3'\beta_0  - \gamma_0 \tilde{X}. $$
Because $F_{\Pi}(\cdot)$ is monotone increasing, we have 
\begin{align*}
& \partial_x P^{11}(Z_1,Z_3,X)/f_V(X) +\partial_x P^{10}(\tilde {Z}_1, \tilde{Z}_3,\tilde X)/f_V(\tilde X) =0  \notag \\
\iff &\alpha_0 + (Z_1-\tilde{Z}_1)'\lambda_0 + (Z_3-\tilde{Z}_3)'\beta_0-\gamma_0 (X - \tilde X)=0
\end{align*}
Note the LHS of the above display is identified from data. Denote $Z_{1,1}$ as the first element of $Z_1$, whose coefficient is set to one. The rest of $Z_1$ is denoted as $Z_{1,-1}$, whose coefficient is denoted as $\lambda_{0,-1}$. Then, we have 
\begin{align*}
\alpha_0 + (Z_{1,-1}-\tilde{Z}_{1,-1})'\lambda_{0,-1} + (Z_3-\tilde{Z}_3)'\beta_0-\gamma_0 (X - \tilde X) = \tilde{Z}_{1,1} - Z_{1,1}. 
\end{align*}
Then, by Assumption \textbf{A4}, we can find $(z_1^{(l)},z_3^{(l)},x^{(l)})_{l = 1}^{d}$ and $(\tilde{z}_1^{(l)},\tilde{z}_3^{(l)},\tilde{x}^{(l)})_{l = 1}^{d}$ such that 
\begin{align*}
\text{rank}\begin{pmatrix}
1 & \cdots & 1\\
z_{1,-1}^{(1)} - \tilde{z}_{1,-1}^{(1)}&  \cdots & z_{1,-1}^{(d)} -\tilde{z}_{1,-1}^{(d)}\\
z_{3}^{(1)} - \tilde{z}_{3}^{(1)}  &  \cdots & z_{3}^{(d)} - \tilde{z}_{3}^{(d)}  \\
x^{(1)} - \tilde{x}^{(1)} &  \cdots & x^{(d)} -\tilde{x}^{(d)}\\
\end{pmatrix} = d.
\end{align*}
Then, we can identify  $(\alpha_0, \lambda_0,\beta_0, \gamma_0)$ by solving the linear system that 
\begin{align*}
\alpha_0 + (z_{1,-1}^{(1)}-\tilde{z}_{1,-1}^{(1)})'\lambda_{0,-1} + (z_3^{(1)}-\tilde{z}_3^{(1)})'\beta_0-\gamma_0 (x^{(1)} - \tilde{x}^{(1)})  = & \tilde{z}_{1,1}^{(1)} - z_{1,1}^{(1)},\\
\vdots &  \\
\alpha_0 + (z_{1,-1}^{(d)}-\tilde{z}_{1,-1}^{(d)})'\lambda_{0,-1} + (z_3^{(d)}-\tilde{z}_3^{(d)})'\beta_0-\gamma_0 (x^{(d)} - \tilde{x}^{(d)})  = & \tilde{z}_{1,1}^{(d)} - z_{1,1}^{(d)}.\\ 
\end{align*}
This concludes the proof.

%In this section we establish the asymptotic theory for each of  the two step  estimators under the conditions when the parameters are regularly identified.
%We first turn attention to the closed form estimator.

%\begin{comment}
\section{Proof of Theorem \ref{thm:aux}}
\setcounter{equation}{0}
\label{sec:auxpf}
For notation simplicity, we write $\tilde{W} = \nu_0 W$, $\tilde{\sigma}_0 = \sigma_0 /\nu_0$, $\tilde{\nu}_0 = 1/\nu_0$, and
\begin{align*}
& Y_2 =  \mathbf{1}\{X \geq \tilde{\nu}_0 \tilde{W} +\eta_2 \}\\
& Y_3 = \tilde{W} + \eta_3 \\
& Y_4 = \tilde{\sigma}_0 \tilde{W} + \eta_4. 
\end{align*}
Because Assumptions B2--B6 hold, by applying \citet[Theorem 1]{SH13} to $Y_3$ and $Y_4$, we can identify the densities for $\nu_0W = \tilde{W}$, $\eta_3$, and $\eta_4$ as well as $\sigma_0/\nu_0 = \tilde{\sigma}_0$.

Then, we have 
\begin{align*}
\partial_{y_3} P(Y_2 = 1,Y_3 \leq y_3|X=x) = & \partial_{y_3} \int F_{\eta_2}(x - \tilde{\nu}_0 w)F_{\eta_3}(y_3 - w)f_{\tilde{W}}(w)dw \\
= & \int F_{\eta_2}(x - \tilde{\nu}_0 w)f_{\eta_3}(y_3 - w)f_{\tilde{W}}(w)dw.
\end{align*}
Applying Fourier transform w.r.t. $y_3$ on both sides, we have 
\begin{align*}
\mathcal{F}(\partial_{y_3} P(Y_2 = 1,Y_3 \leq \cdot|X=x))(t) = \mathcal{F}(F_{\eta_2}(x - \tilde{\nu}_0 \cdot)f_{\tilde{W}}(\cdot))(t)\mathcal{F}(f_{\eta_3}(\cdot))(t),
\end{align*}
where for a generic function $g(w)$, 
\begin{align*}
\mathcal{F}(g(\cdot))(t) = \frac{1}{\sqrt{2\pi}}\int \exp(-2\pi itw)g(w)dw. 
\end{align*}
Therefore, 
\begin{align}
\frac{\mathcal{F}^{-1}\left(\frac{\mathcal{F}(\partial_{y_3} P(Y_2 = 1,Y_3 \leq \cdot|X=x))(\cdot)}{\mathcal{F}(f_{\eta_3}(\cdot))(\cdot)}\right)(w)}{f_{\tilde{W}}(w)} = F_{\eta_2}(x - \tilde{\nu}_0 w),
\label{eq:1}
\end{align}
where for a generic function $g(w)$, 
\begin{align*}
\mathcal{F}^{-1}(g(\cdot))(t) = \frac{1}{\sqrt{2\pi}} \int \exp(2\pi itw)g(w)dw. 
\end{align*}
Note the LHS of \eqref{eq:1} can be identified from data. We choose two pairs $(x,w)$ and $(x',w')$ such that   $w \neq w'$ and 
\begin{align*}
\frac{\mathcal{F}^{-1}\left(\frac{\mathcal{F}(\partial_{y_3} P(Y_2 = 1,Y_3 \leq \cdot|X=x))(\cdot)}{\mathcal{F}(f_{\eta_3}(\cdot))(\cdot)}\right)(w)}{f_{\tilde{W}}(w)} = \frac{\mathcal{F}^{-1}\left(\frac{\mathcal{F}(\partial_{y_3} P(Y_2 = 1,Y_3 \leq \cdot|X=x'))(\cdot)}{\mathcal{F}(f_{\eta_3}(\cdot))(\cdot)}\right)(w')}{f_{\tilde{W}}(w')}.
\end{align*}
Then, given the monotonicity of $F_{\eta_2}$, we have 
\begin{align*}
x - \tilde{\nu}_0 w = x' - \tilde{\nu}_0 w',
\end{align*}
or 
\begin{align*}
\tilde{\nu}_0 = (x-x')/(w-w'), 
\end{align*}
which is identified. Given the identification of $\tilde{\nu}_0$ and the distribution of $\tilde{W}$, we can identify the distribution of $W= \tilde{\nu}_0 \tilde{W}$. Recall $F_{\eta_1}(\cdot)$ and $f_{\eta_2}(\cdot)$ are the CDF and PDF of $\eta_1$ and $\eta_2$, respectively. Then, we have 
\begin{align*}
P(Y_2=1|X=x) = P(W+\eta_2 \leq x).
\end{align*}
Because $X$ has full support, we can identify the distribution of $W+\eta_2$. Then, it follows from standard deconvolution argument and the fact that the distribution of $W$ is identified that we can identify the distribution of $\eta_2$. In addition, note that 
\begin{align*}
P^{11}(z_1,z_3,x) = & P(Y_1=1,Y_2=1|Z_1=z_1,Z_3=z_3,X=x) \\
=  & \int F_{\eta_1}(z_1'\lambda_0 + z_3'\beta_0+\alpha_0-\gamma_0 w)F_{\eta_2}(x-w)f_W(w)dw
\end{align*}
and 
\begin{align*}
P^{10}(z_1,z_3,x) = & P(Y_1=1,Y_2=0|Z_1=z_1,Z_3=z_3,X=x) \\
= & \int F_{\eta_1}(z_1'\lambda_0 + z_3'\beta_0-\gamma_0 w)(1-F_{\eta_2}(x-w))f_W(w)dw.
\end{align*}
Taking derivatives of $P^{11}(z_1,z_3,x)$ and $P^{10}(z_1,z_3,x)$ w.r.t. $x$, we have 
\begin{align}
\label{eq:p11''}
\partial_x P^{11}(z_1,z_3,x) = \int F_{\eta_1}(z_1'\lambda_0 + z_3'\beta_0+\alpha_0-w)f_{\eta_2}(x- w)f_W(w)dw
\end{align}
and 
\begin{align}
\label{eq:p10''}
-\partial_x P^{10}(z_1,z_3,x) = \int F_{\eta_1}(z_1'\lambda_0 + z_3'\beta_0-\gamma_0w)f_{\eta_2}(x- w)f_W(w)dw.
\end{align}

Applying Fourier transform on both sides of \eqref{eq:p11''} and \eqref{eq:p10''}, we have 
\begin{align}
\label{eq:f11'}
\mathcal{F}(\partial_x P^{11}(z_1,z_3,\cdot)) = \mathcal{F}( F_{\eta_1}(z_1'\lambda_0 + z_3'\beta_0+\alpha_0- \cdot)f_W(\cdot))\mathcal{F}(f_{\eta_2}(\cdot))
\end{align}
and
\begin{align*}
%\label{eq:f10'}
\mathcal{F}(-\partial_x P^{10}(z_1,z_3,\cdot)) = \mathcal{F}( F_{\eta_1}(z_1'\lambda_0 + z_3'\beta_0-\cdot)f_W(\cdot))\mathcal{F}(f_{\eta_2}(\cdot)).
\end{align*}

Then, by \eqref{eq:f11'}, we can identify $F_{\eta_1}(z_1'\lambda_0 + z_3'\beta_0+\alpha_0 - \cdot)$ by 
\begin{align*}
F_{\eta_1}(z_1'\lambda_0 + z_3'\beta_0+\alpha_0 - \gamma_0 \cdot) = \mathcal{F}^{-1}\left(\frac{\mathcal{F}(\partial_x P^{11}(z_1,z_3,\cdot))}{\mathcal{F}(f_{\eta_2}(\cdot))} \right)(\cdot)/f_W(\cdot).
\end{align*}
Similarly, we can identify 
\begin{align*}
F_{\eta_1}(z_1'\lambda_0 + z_3'\beta_0 - \gamma_0 \cdot) = \mathcal{F}^{-1}\left(\frac{\mathcal{F}(-\partial_x P^{10}(z_1,z_3,\cdot))}{\mathcal{F}(f_{\eta_2}(\cdot))} \right)(\cdot)/f_W(\cdot).
\end{align*}

Because $F_{\eta_1}(\cdot)$ is monotone increasing, we have 
\begin{align*}
& \mathcal{F}^{-1}\left(\frac{\mathcal{F}(\partial_x P^{11}(z_1,z_3,\cdot))}{\mathcal{F}(f_{\eta_2}(\cdot))} \right)(w)/f_W(w) =\mathcal{F}^{-1}\left(\frac{\mathcal{F}(-\partial_x P^{10}(\tilde{z}_1,\tilde{z}_3,\cdot))}{\mathcal{F}(f_{\eta_2}(\cdot))} \right)(\tilde{w})/f_W(\tilde{w})  \notag \\
\iff &\alpha_0 + (z_1-\tilde{z}_1)'\lambda_0 + (z_3-\tilde{z}_3)'\beta_0-\gamma_0 (w - \tilde w)=0
\end{align*}

Then, by Assumption B7, we can find $(z_1^{(l)},z_3^{(l)},w^{(l)})_{l = 1}^{d}$ and $(\tilde{z}_1^{(l)},\tilde{z}_3^{(l)},\tilde{w}^{(l)})_{l = 1}^{d}$ such that 
\begin{align*}
\text{rank}\begin{pmatrix}
1 & \cdots & 1\\
z_{1,-1}^{(1)} - \tilde{z}_{1,-1}^{(1)}&  \cdots & z_{1,-1}^{(d)} -\tilde{z}_{1,-1}^{(d)}\\
z_{3}^{(1)} - \tilde{z}_{3}^{(1)}  &  \cdots & z_{3}^{(d)} - \tilde{z}_{3}^{(d)}  \\
w^{(1)} - \tilde{w}^{(1)} &  \cdots & w^{(d)} -\tilde{w}^{(d)}\\
\end{pmatrix} = d.
\end{align*}
Then, we can identify  $(\alpha_0, \lambda_0,\beta_0, \gamma_0)$ by solving the linear system that 
\begin{align*}
\alpha_0 + (z_{1,-1}^{(1)}-\tilde{z}_{1,-1}^{(1)})'\lambda_{0,-1} + (z_3^{(1)}-\tilde{z}_3^{(1)})'\beta_0-\gamma_0 (w^{(1)} - \tilde{w}^{(1)})  = & \tilde{z}_{1,1}^{(1)} - z_{1,1}^{(1)},\\
\vdots &  \\
\alpha_0 + (z_{1,-1}^{(d)}-\tilde{z}_{1,-1}^{(d)})'\lambda_{0,-1} + (z_3^{(d)}-\tilde{z}_3^{(d)})'\beta_0-\gamma_0 (w^{(d)} - \tilde{w}^{(d)})  = & \tilde{z}_{1,1}^{(d)} - z_{1,1}^{(d)}.\\ 
\end{align*}
This concludes the proof.

%By finding the two pairs $((x_1,w), (x_1',w'))$ and $((\tilde{x}_1,\tilde{w}), (\tilde{x}_1',\tilde{w}'))$ such that $w - w' \neq \tilde{w} - \tilde{w}'$, 
%\begin{align*}
%F_{\eta_1}(x_1+\alpha_0- \gamma_0 w) = F_{\eta_1}(x_1'- \gamma_0 w'), \quad \text{and} \quad  F_{\eta_1}(\tilde{x}_1+\alpha_0- \gamma_0 \tilde{w}) = F_{\eta_1}(\tilde{x}_1'- \gamma_0 \tilde{w}') 
%\end{align*}
%we can identify both $\alpha_0$ and $\gamma_0$ as the solution of the following linear system: 
%\begin{align*}
%& \alpha_0+ \gamma_0 (w'-w) = x_1'- x_1 
%& \alpha_0 +  \gamma_0 (\tilde{w}'-\tilde{w}) = \tilde{x}_1'- \tilde{x}_1 .
%\end{align*}

\section{Finite Sample Properties} 
\label{sec:sim}
\setcounter{equation}{0}
In this section we explore the finite sample properties of the proposed estimation procedure via a simulation study. % In it we aim to explore how well identified the parameter of interest when the second instrumental variable is not present in the model. 
We will also see how sensitive the performance of the proposed estimator 
is to the factor structure assumption. As a base comparison, we also report results for the estimator proposed in \cite{vytlacilyildiz} to see how sensitive it is 
to their second instrument restriction.

Our data are simulated from base models of the form 
\begin{align}
Y_1 &= {\bf 1}\{X_1 + \alpha_0 Y_2 - U \geq 0 \} \label{eq:y1}\\
Y_2 &= {\bf 1}\{X - V>0 \} \label{eq:y2},
\end{align}
where $X_1$ is binary with success probability 0.6, $X$ has marginal distribution $\N(0,1)$, $X_1$ and $X$ are mutually independent, $(X_1,X) \perp (V,\Pi)$, $U = \gamma_0V + \Pi $. $(V,\Pi)$ are distributed independently of each other, where $\Pi$ is distributed following a standard normal distribution, and $V$ is distributed either standard normal, Laplace, or $T(3)$.
The parameters  $(\alpha_0,\gamma_0) = (-0.25,1.2)$ or $(0.5,1.2)$. 

\medskip
Since $X_1$ is discrete, \citeauthor{vytlacilyildiz}'s (\citeyear{vytlacilyildiz}) identification condition does not hold. However, the identification condition in this paper becomes
$$|\alpha| \leq \mbox{length of the support of }X,$$
which holds. 

\medskip
For each choice of sample size $n = 100,200,400,800,1,600$, we simulate 280 samples and report the bias, standard deviation (std), root mean squared error (RMSE), and median absolute deviation (MAD) for both \citeauthor{vytlacilyildiz}'s (\citeyear{vytlacilyildiz}) estimator (VY) and ours (KMZ). For implementation, we use the second order local polynomial along with Gaussian kernels to nonparametrically estimate the $\partial_2 P^{11}(x_1,x)$ and $\partial_2 P^{10}(x_1,x)$. The bandwidth we use is $h_1 = \sigma_xN^{-1/7}$ where $\sigma_x$ is the standard deviation of $X$. $f_V(x)$ is nonparametrically estimated using a local linear estimator with the tuning parameter $h_2 = \sigma_x N^{-1/6}$. %Recent work by \cite{liparmeter} discuss bandwidth selection methods for estimating derivatives of regression functions which could prove useful for our estimator at hand.

\medskip
As results from the table indicate, the finite sample performance of our estimator generally agrees with the asymptotic theory. The RMSE for the estimator proposed here is decreasing as the sample size increases, as one could expect given the consistency property of our estimator. Besides, the decay rate of the RMSE and MAD is about $\sqrt{2}$ when $n \geq 400$ as sample sizes doubles, in line with the parametric rate of convergence of our estimator. 

\medskip
\citeauthor{vytlacilyildiz}'s (\citeyear{vytlacilyildiz}) estimator,  which does not exploit the factor structure, demonstrates inconsistency for certain parameter values, as indicated by the bias and median bias not shrinking with the sample size. In addition, the RMSE and MAD do not appear to decline at all, which also suggests that \citeauthor{vytlacilyildiz}'s (\citeyear{vytlacilyildiz}) estimator is inconsistent in these designs.\footnote{Because $X_1$ is binary, \citeauthor{vytlacilyildiz}'s (\citeyear{vytlacilyildiz}) estimator can only take 3 possible values: 0, -1 or 1. In particular, when $\alpha=0.5$, in most of the replications, the estimator takes values 0 or 1. When $\alpha=-0.25$, in most of the replications, the estimator takes value -1. In both of these cases, the MAD remains constant over the different sample sizes.} 

\begin{table}[H]
	\centering
	\caption{Normal $V$, $\alpha = 0.5$}
	\begin{adjustbox}{width=1\textwidth}
		\begin{tabular}{r|rrr|rrr|rrr|rrr|rrr|rrr} \hline \hline
			\multicolumn{1}{c|}{$\Pi$}	& \multicolumn{6}{c|}{Normal}             & \multicolumn{6}{c|}{Laplace}            & \multicolumn{6}{c}{T(3)} \\ \hline
			& \multicolumn{3}{c|}{kmz} & \multicolumn{3}{c|}{vy} & \multicolumn{3}{c|}{kmz} & \multicolumn{3}{c|}{vy} & \multicolumn{3}{c|}{kmz} & \multicolumn{3}{c}{vy} \\ \hline
			\multicolumn{1}{c|}{N} & \multicolumn{1}{l}{Bias } & \multicolumn{1}{l}{RMSE} & \multicolumn{1}{l|}{MAD} & \multicolumn{1}{l}{Bias } & \multicolumn{1}{l}{RMSE} & \multicolumn{1}{l|}{MAD} & \multicolumn{1}{l}{Bias } & \multicolumn{1}{l}{RMSE} & \multicolumn{1}{l|}{MAD} & \multicolumn{1}{l}{Bias } & \multicolumn{1}{l}{RMSE} & \multicolumn{1}{l|}{MAD} & \multicolumn{1}{l}{Bias } & \multicolumn{1}{l}{RMSE} & \multicolumn{1}{l|}{MAD} & \multicolumn{1}{l}{Bias } & \multicolumn{1}{l}{RMSE} & \multicolumn{1}{l}{MAD} \\ \hline
			100   & -0.026 & 0.665 & 0.660 & -0.246 & 0.658 & 0.500 & 0.032 & 0.634 & 0.560 & -0.293 & 0.658 & 0.500 & 0.010 & 0.676 & 0.665 & -0.225 & 0.662 & 0.500 \\
			200   & 0.004 & 0.591 & 0.475 & -0.329 & 0.633 & 0.500 & -0.015 & 0.568 & 0.400 & -0.336 & 0.612 & 0.500 & -0.003 & 0.616 & 0.495 & -0.279 & 0.629 & 0.500 \\
			400   & 0.005 & 0.483 & 0.365 & -0.341 & 0.573 & 0.500 & 0.030 & 0.459 & 0.310 & -0.323 & 0.559 & 0.500 & 0.018 & 0.542 & 0.405 & -0.314 & 0.589 & 0.500 \\
			800   & 0.065 & 0.456 & 0.300 & -0.348 & 0.544 & 0.500 & 0.096 & 0.391 & 0.250 & -0.357 & 0.511 & 0.500 & 0.046 & 0.462 & 0.295 & -0.346 & 0.552 & 0.500 \\
			1,600  & 0.040 & 0.321 & 0.195 & -0.413 & 0.503 & 0.500 & 0.017 & 0.294 & 0.190 & -0.450 & 0.506 & 0.500 & 0.034 & 0.371 & 0.240 & -0.368 & 0.506 & 0.500 \\
		\end{tabular}
	\end{adjustbox}
	\label{tab:1}%
\end{table}%

\begin{table}[H]
	\centering
	\caption{Normal $V$, $\alpha = -0.25$}
	\begin{adjustbox}{width=1\textwidth}
		\begin{tabular}{r|rrr|rrr|rrr|rrr|rrr|rrr} \hline \hline
			\multicolumn{1}{c|}{$\Pi$}	& \multicolumn{6}{c|}{Normal}             & \multicolumn{6}{c|}{Laplace}            & \multicolumn{6}{c}{T(3)} \\ \hline
			& \multicolumn{3}{c|}{kmz} & \multicolumn{3}{c|}{vy} & \multicolumn{3}{c|}{kmz} & \multicolumn{3}{c|}{vy} & \multicolumn{3}{c|}{kmz} & \multicolumn{3}{c}{vy} \\ \hline
			\multicolumn{1}{c|}{N} & \multicolumn{1}{l}{Bias } & \multicolumn{1}{l}{RMSE} & \multicolumn{1}{l|}{MAD} & \multicolumn{1}{l}{Bias } & \multicolumn{1}{l}{RMSE} & \multicolumn{1}{l|}{MAD} & \multicolumn{1}{l}{Bias } & \multicolumn{1}{l}{RMSE} & \multicolumn{1}{l|}{MAD} & \multicolumn{1}{l}{Bias } & \multicolumn{1}{l}{RMSE} & \multicolumn{1}{l|}{MAD} & \multicolumn{1}{l}{Bias } & \multicolumn{1}{l}{RMSE} & \multicolumn{1}{l|}{MAD} & \multicolumn{1}{l}{Bias } & \multicolumn{1}{l}{RMSE} & \multicolumn{1}{l}{MAD} \\ \hline
			
			100   & -0.088 & 0.650 & 0.555 & -0.466 & 0.710 & 0.750 & 0.092 & 0.614 & 0.530 & -0.358 & 0.650 & 0.750 & 0.004 & 0.619 & 0.505 & -0.430 & 0.681 & 0.750 \\
			200   & -0.035 & 0.599 & 0.420 & -0.446 & 0.681 & 0.750 & 0.012 & 0.552 & 0.385 & -0.485 & 0.689 & 0.750 & -0.008 & 0.583 & 0.425 & -0.463 & 0.687 & 0.750 \\
			400   & -0.016 & 0.467 & 0.325 & -0.487 & 0.668 & 0.750 & -0.010 & 0.388 & 0.200 & -0.552 & 0.686 & 0.750 & -0.003 & 0.496 & 0.340 & -0.489 & 0.675 & 0.750 \\
			800   & -0.028 & 0.324 & 0.165 & -0.591 & 0.697 & 0.750 & 0.006 & 0.279 & 0.180 & -0.599 & 0.701 & 0.750 & 0.032 & 0.399 & 0.230 & -0.533 & 0.682 & 0.750 \\
			1,600  & -0.006 & 0.244 & 0.150 & -0.654 & 0.718 & 0.750 & -0.028 & 0.204 & 0.130 & -0.714 & 0.738 & 0.750 & -0.021 & 0.279 & 0.190 & -0.629 & 0.710 & 0.750 \\
		\end{tabular}%
	\end{adjustbox}
	\label{tab:1'}%
\end{table}%

In the following designs we also consider three DGPs (DGPs 1--3) such that the one-factor model does not hold but the identification assumption in \cite{vytlacilyildiz} does. In this case, our simulation results show that while, as expected, the estimator VY is still valid, our estimator still performs reasonably well. Interestingly, this offers suggestive evidence that our estimator is robust to some degree of misspecification. As such, these results complement previous work highlighting the robustness of rank type estimators to misspecification - see \cite{khantamerjbes}. In DGP 4, the identification assumptions in both \cite{vytlacilyildiz} and our paper hold. In this case, we found that our estimator has similar performance as that proposed by \cite{vytlacilyildiz}. 

The outcome and selection equations are the same as \eqref{eq:y1} and \eqref{eq:y2}, respectively. Then, 
\begin{enumerate}
	\item[DGP 1]: $(X_1,X)$ is jointly standard normally distributed. Let $(e_1,e_2)$ jointly Laplace distributed with mean zero and variance-covariance matrix $\Sigma = \begin{pmatrix}
	1 & -0.5 \\
	-0.5 & 1
	\end{pmatrix}$, $e_3$ and $e_4$ are uniformly distributed on $(0,1)$, independent of each other, and independent of $(e_1,e_2)$, $V = e_1+e_3-0.5$, $U = e_2+e_4-0.5$, and $\alpha = -0.25$.  
	\item[DGP 2]: $(X_1,X)$ are the same as above, $U= e_1 + e_2-0.5$, and $V= e_1 + e_3-0.5$, where $e_1$ is standard normally distributed, $(e_2,e_3)$ are uniformly distributed on $(0,1)$, $(e_1,e_2,e_3)$ are mutually independent, and $\alpha = -0.25$.  
	\item[DGP 3]:  $(X_1,X)$ are the same as above, $V = \frac{\exp(e_1+e_2-0.5)-1}{4}$, $U = \frac{\exp(e_1+e_3-0.5)-1}{4}$, $(e_1,e_2,e_3)$ are defined as above, and $\alpha = -0.5$. 
	\item[DGP 4]:  $(X_1,X)$ are the same as above, $V$ is  Laplace distributed with mean zero and standard derivation $0.5$, $U = V + V'-0.5$, where $V'$ is uniform distributed on $(0,1)$ and is independent of $V$, and $\alpha = -0.25$. 	
\end{enumerate}
For DGPs 1, 2, and 4, when computing $\partial_2 P^{11}(x_1,x)$ and $\partial_2 P^{10}(x_1,x)$, we use bandwidths $h_1 = \sigma_{x1}N^{-1/7}$ and $h = \sigma_{x}N^{-1/7}$ for variables $X_1$ and $X$, respectively, where $\sigma_{x1}$ and $\sigma_{x}$ are the standard errors of $X_1$ and $X$, respectively. To estimate the density $f_V(x)$, we use bandwidth $h_2 = \sigma_x N^{-1/6}$. For DGP 3, we use $h_1 =h_2 = h = \sigma_{x1}N^{-1/5}$. In all simulations, we use 280 replications.

\begin{table}[H]
	\centering
	\caption{Alternative DGPs}
	\begin{adjustbox}{width=1\textwidth}
		\begin{tabular}{r|rrr|rrr|rrr|rrr} \hline \hline 
			& \multicolumn{6}{c|}{DGP 1}                     & \multicolumn{6}{c}{DGP 2} \\ \hline 
			& \multicolumn{3}{c|}{kmz} & \multicolumn{3}{c|}{vy} & \multicolumn{3}{c|}{kmz} & \multicolumn{3}{c}{vy} \\ \hline 
			\multicolumn{1}{l|}{N} & \multicolumn{1}{l}{Bias } & \multicolumn{1}{l}{RMSE} & \multicolumn{1}{l|}{MAD} & \multicolumn{1}{l}{Bias } & \multicolumn{1}{l}{RMSE} & \multicolumn{1}{l|}{MAD} & \multicolumn{1}{l}{Bias } & \multicolumn{1}{l}{RMSE} & \multicolumn{1}{l|}{MAD} & \multicolumn{1}{l}{Bias } & \multicolumn{1}{l}{RMSE} & \multicolumn{1}{l}{MAD} \\ \hline 
			
			100	&-0.065 &	0.678&	0.600&	-0.055&	0.666&	0.535&	-0.058&	0.621&	0.505&	-0.05&	 0.621&	0.470\\
			200	&-0.118&	0.543&	0.370&	-0.080& 	0.497&	0.320&	-0.122&	0.523&	0.350&	-0.097&	0.495&	0.350\\
			400	&-0.117&	0.413&	0.280&	-0.071&	0.378&	0.245&	-0.062&	0.335&	0.215&	-0.033&	0.316&	0.220\\
			800	&-0.102&	0.287&	0.170&	-0.062&	0.243&	0.160&	-0.031&	0.242&	0.150&	-0.008&	0.215&	0.150\\
			1,600&	-0.071&	0.193&	0.140&	-0.035&	0.155&	0.100&	-0.038&	0.167&	0.100&	-0.031&	0.158&	0.100\\\hline 
			& \multicolumn{6}{c|}{DGP 3}                     & \multicolumn{6}{c}{DGP 4} \\ \hline 
			100	&-0.012&	0.583&	0.480&	-0.015&	0.565&	0.430&	-0.057&	0.401&	0.240&	-0.066&	0.422&	0.240\\
			200	&-0.061&	0.425&	0.275&	-0.068&	0.399&	0.270&	-0.041&	0.282&	0.180&	-0.049&	0.263&	0.145\\
			400	&-0.041&	0.259&	0.170&	-0.042&	0.237&	0.155&	-0.062&	0.184&	0.135&	-0.047&	0.186&	0.120\\
			800	&-0.061&	0.219&	0.140&	-0.047&	0.182&	0.120&	-0.029&	0.119&	0.080&	-0.034&	0.115&	0.070\\
			1,600&	-0.038&	0.130&	0.080&	-0.035&	0.119&	0.080&	-0.024&	0.090&	0.060&	-0.022&	0.086&	0.070\\ \hline	
		\end{tabular}%
	\end{adjustbox}
	\label{tab:alter}%
\end{table}%

In the first three DGPs, we see that VY's estimator has better performance in terms of both bias and MSE. On the other hand, although the models do not have a factor structure, our estimator still performs reasonably well. In the last DGP, support conditions in both \cite{vytlacilyildiz} and our paper hold. Table \ref{tab:alter} shows that our and \citeauthor{vytlacilyildiz}'s (\citeyear{vytlacilyildiz}) estimators have similar performance in terms of bias and MSE. Although our estimator is expected to be more efficient as we use the factor structure in estimation, it is not. We conjecture that it is because our estimator does not necessarily use all the information, or in other words, achieve the semiparametric efficiency bound. To establish the semiparametric efficient estimator in the model with and without the factor structure is an interesting yet challenging task. We leave it as a topic for future research.

\section{Identification with and without Factor Structure}
\label{sec:vys}
\setcounter{equation}{0}
\subsection{Identification Without Auxiliary Measurements}
\label{sec:vys_nomeas}
In this section, we discuss the information content of factor structure. For illustration purpose, we focus on the ``condensed" model:
\begin{equation}
\begin{aligned}
& Y_1 = \mathbf{1}\{X_1 + \alpha_0 Y_2 - U \geq 0 \} \\
& Y_2 = \mathbf{1}\{X - V \geq 0 \}.
\end{aligned}
\label{eq:11}
\end{equation}
\begin{ass} \mbox{ } \\
	1. $(X_1,X) \perp (U,V)$. \\
	2. $(X_1,X)$ are continuously distributed with absolute continuous joint density w.r.t. Lebesgue measure. The conditional support of $X_1$ given $X$ is $[a,b]$.\\
	3. $V$ is continuously distributed over $\Re$ and its density w.r.t. Lebesgue measure exist.
	\label{ass:indep}
\end{ass}

\begin{theorem}
	If	Assumption \ref{ass:indep} holds, then $|\alpha_0| \leq b-a$ is necessary and sufficient for $\alpha_0$ to be identified. 
	\label{thm:id_suf}
\end{theorem}

We note that under Assumption \ref{ass:indep}, $|\alpha_0| \leq b-a$ is equivalent to the fact that we can find $x_1$ and $\tilde{x}_1$ in the support of $X_1$ such that $\alpha_0 = x_1 - \tilde{x}_1$. 

Next, we assume, in addition to Assumption \ref{ass:indep}, the factor structure, i.e., \eqref{eq:factor1} in Section \ref{tmfs}. 
Our rank estimator can be written as an M-estimator
\begin{equation*}
\hat{\theta} = \arg\max_\theta Q_{n}(\theta) \equiv \sum_{i \neq j}\hat{g}_{i,j}(\theta)\end{equation*}
in which
\begin{eqnarray*}
	\hat{g}_{i,j}(\theta) &= & [\mathbf{1}\{\partial_2 \hat{P}^{11}(X_{1,i},X_i)/\hat{f}_V(X_i) + \partial_2 \hat{P}^{10}(X_{1,j},X_j)/\hat{f}_V(X_j)\geq 0 \}\mathbf{1}\{\Phi(X_{1,i},X_i,X_{1,j},X_j;\theta) \geq 0\} \\
	& +& \mathbf{1}\{\partial_2 \hat{P}^{11}(X_{1,i},X_i)/\hat{f}_V(X_i) + \partial_2 \hat{P}^{10}(X_{1,j},X_j)/\hat{f}_V(X_j) <0\}\mathbf{1}\{\Phi(X_{1,i},X_i,X_{1,j},X_j;\theta) <0 \} ], 
\end{eqnarray*}
with 
\[\Phi(x_1,x,\tilde{x}_1,\tilde{x};\theta) = x_1 + \alpha - \gamma x - (\tilde{x}_1 - \gamma \tilde{x}).\]
We will study the asymptotic properties of this estimator in Section \ref{asymprop}.

The information content explored by the M-estimator can be summarized as follows: 
\begin{equation*}
\begin{aligned}
\mathcal{A}_2(\theta) = \{ (X_1,\tilde{X}_1,X,\tilde{X}), &\Phi(X_1,X,\tilde{X}_1,\tilde{X};\theta_0) \geq 0 > \Phi(X_1,X,\tilde{X}_1,\tilde{X};\theta) \\
& \mbox{ or }  \Phi(X_1,X,\tilde{X}_1,\tilde{X};\theta_0) < 0 \leq \Phi(X_1,X,\tilde{X}_1,\tilde{X};\theta)\}.
\end{aligned}
\end{equation*}
Then we cannot distinguish, from the true parameter $\theta_0$, all impostors in 
$$\overline{\mathcal{A}}_2 = \{\theta: P(\mathcal{A}_2(\theta)) = 0 \}.$$
In the condensed model, if $\Supp(X_1,X) = [a,b] \times [c,d]$, then $\theta_0$ is identified if $|\alpha_0| < b-a + |\gamma_0|(d-c)$. Recall Theorem \ref{thm:id_suf}, without imposing factor structure, the necessary and sufficient condition for achieving identification is $|\alpha_0| \leq b-a$. Therefore, the blue area in the Figure below is the additional parts of parameter space that are identified with factor structure but not otherwise. 
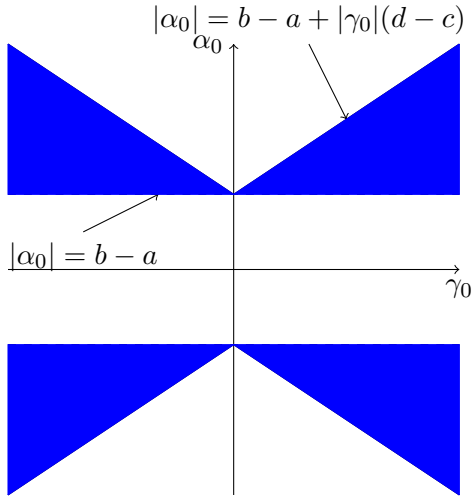
\begin{figure}[H]
	\centering
	\begin{tikzpicture}
	\coordinate (y) at (0,3);
	\coordinate (x) at (3,0);
	%	\node [below right] at (2,0) {$a$};
	%	\node [below right] at (3.5, 0) {$b$};
	%	\node [above left] at (0,2.5) {$a + \alpha_0$};
	%	\node [above left] at (0,4) {$b + \alpha_0$};  
	%	\node [below left] at (0,2) {$a$ };
	%	\node [below left] at (0,3.5) {$b$};  	
	\draw[<-] (y) node[left] {$\alpha_0$} -- (0,-3); 
	\draw[->] (-3,0) --  (x) node[below]{$\gamma_0$};
	\draw [dashed] (-3,-1) -- (3,-1);
	\draw [dashed] (-3,1) -- (3,1);
	\draw  (-3,3) -- (0,1);
	\draw  (3,3) -- (0,1);
	\draw  (-3,-3) -- (0,-1);
	\draw  (3,-3) -- (0,-1);
	\node[above] at (1,3)  {$|\alpha_0| = b-a + |\gamma_0|(d-c)$};	    
	\path[->] (1,3) edge  (1.5,2);
	\node[below] at (-2,0.5)  {$|\alpha_0| = b-a$};	    
	\path[->] (-2,0.5) edge  (-1,1);
	\draw [blue, fill=blue] (-3,3) -- (0,1) -- (-3,1) -- (-3,3);
	\draw [blue, fill=blue] (3,3) -- (0,1) -- (3,1) -- (3,3);
	\draw [blue, fill=blue] (-3,-3) -- (0,-1) -- (-3,-1) -- (-3,-3);
	\draw [blue, fill=blue] (3,-3) -- (0,-1) -- (3,-1) -- (3,-3);	
	\end{tikzpicture}
	\caption{Identifying Power of Factor Structure}
	\label{fig:2}
\end{figure}

\begin{theorem}	
	Assumption \ref{ass:indep} holds. When $|\alpha_0|> b-a$, the sharp identified set for $\alpha_0$ is 
	$$\mathcal{A}^* = \{\alpha: \alpha > b-a \mbox{ if } \alpha_0 > 0 \mbox{ and } \alpha < a-b \mbox{ if } \alpha_0 < 0  \}.$$ 
	\label{thm:partialid}
\end{theorem}

Theorem \ref{thm:partialid} highlights that, in the case without the factor structure and $\alpha_0$ does not satisfy the parameter restriction, except for the fact that the sign of $\alpha_0$ is identified, we actually cannot say much about the value of $|\alpha_0|$. When we assume the factor structure, the parameter is still not identified if $|\alpha_0| > b-a + |\gamma_0|(d-c)$. In addition, suppose $\alpha_0>0$. In this case, if we do not impose factor structure, by Theorem \ref{thm:partialid}, the sharp identified set is $\{\alpha:\alpha > b-a\}$ while with the factor structure, the identified set (not necessarily sharp) is $\alpha > b-a + |\gamma|(d-c)$. This implies, when identification fails in both cases, the  blue area is also the extra identifying power on the identified set given by the factor structure. 

\subsection{Identification with two auxiliary measurements}
\label{sec:indep2}
Next, we expand our condensed model to include two continuous measurements. We show in this case, without the factor structure, $\alpha_0$ is not identified. This is in contrast with the identification result established in Theorem \ref{thm:aux}. 

Suppose in addition to \eqref{eq:11}, we also observe two continuous measurements of $W$ denoted as $Y_3$ and $Y_4$. One example of such $Y_3$ and $Y_4$ are described in \eqref{eq:measurements}. 

\begin{ass} \mbox{ } \\
	1. $(X_1,X) \perp (U,V,Y_3,Y_4)$. \\
	2. $(X_1,X)$ are continuously distributed with absolute continuous joint density w.r.t. Lebesgue measure. The conditional support of $X_1$ given $X$ is $[a,b]$.\\
	3. $V$ is continuously distributed over $\Re$ and its density w.r.t. Lebesgue measure exist.
	\label{ass:indep2}
\end{ass}

\begin{theorem}
	If	Assumption \ref{ass:indep2} holds, then $|\alpha_0| \leq b-a$ is necessary and sufficient for $\alpha_0$ to be identified. 
	\label{thm:id_suf2}
\end{theorem}

The proof of Theorem \ref{thm:id_suf2} is similar to that of Theorem \ref{thm:id_suf}, and thus, is omitted. In the proof of Theorem \ref{thm:id_suf}, we show that when $|\alpha_0| > b-a$, we can find an impostor $\alpha \neq \alpha_0$ and $\tilde{U}$ such that for any $x_1 \in [a,b]$ and any $v \in \Supp(V)$, we have
\begin{align*}
&P(\tilde{U} \leq x_1 + \alpha|V=v) = P(U \leq x_1 + \alpha_0|V=v)\\
&P(\tilde{U} \leq x_1|V=v) = P(U \leq x_1|V=v).
\end{align*}
This implies the conditional CDF of $(Y_1,Y_2)$ given $(X_1,X)$ under the DGPs $(U,V,\alpha_0)$ and $(\tilde{U},V,\alpha)$ are the same, and thus, $\alpha_0$ is observationally equivalent to the impostor $\alpha$. Similarly, with the two continuous measurements, we can use the exact same construction of $\tilde{U}$ and $\alpha$ to show that, for any $x_1 \in [a,b]$ and $(v,y_3,y_4)\in \Supp(V,Y_3,Y_4)$, we have 
\begin{align*}
&P(\tilde{U} \leq x_1 + \alpha|V=v,Y_3=y_3,Y_4=y_4) = P(U \leq x_1 + \alpha_0|V=v,Y_3=y_3,Y_4=y_4)\\
&P(\tilde{U} \leq x_1|V=v,Y_3=y_3,Y_4=y_4) = P(U \leq x_1|V=v,Y_3=y_3,Y_4=y_4).
\end{align*}
This implies the conditional CDF of $(Y_1,Y_2,Y_3,Y_4)$ given $(X_1,X)$ under the DGPs $(U,V,Y_3,Y_4,\alpha_0)$ and $(\tilde{U},V,Y_3,Y_4,\alpha)$ are the same too. Such non-identification result holds even when $X$ has full support. 

\section{Proof of Theorem \ref{thm:id_suf}}
\setcounter{equation}{0}
\label{sec:sur_proof}
Denote $P^{ij}(x_1,x) = Prob(Y_1 = i,Y_2 = j|X_1 =x_1,X=x)$. Then 
\begin{equation}
\begin{aligned}
P^{11}(x_1,x) & = \int_{-\infty}^{x}F_U(x_1 + \alpha_0|V=v)f(v)dv \\
P^{10}(\tilde{x}_1,x) & = \int^{+\infty}_{x}F_U(\tilde{x}_1|V=v)f(v)dv.
\label{eq:without_P}
\end{aligned}
\end{equation}
Taking derivatives w.r.t. the second argument of the the LHS function, we have 
\begin{equation*}
\begin{aligned}
\partial_2 P^{11}(x_1,x) & = F_U(x_1 + \alpha_0|V=x)f(x) \\
\partial_2 P^{10}(\tilde{x}_1,x) & = -F_U(\tilde{x}_1|V=x)f(x).
\end{aligned}
\end{equation*}
If $|\alpha_0| \leq b-a$, then there exists a pair $(x_1,\tilde{x}_1)$ such that $x_1 + \alpha_0 = \tilde{x}_1$. This pair can be identified by checking the equation below:
$$\partial_2 P^{11}(x_1,x)/f(x) + \partial_2 P^{10}(\tilde{x}_1,x)/f(x) = 0.$$ 
This concludes the sufficient part. 

\medskip
When $\alpha_0 < a-b$, for any $\alpha < \alpha_0$, we can define 
\begin{align*}
& \tilde{U} = U + \alpha- \alpha_0 & \mbox{ if } & & U \leq b + \alpha_0 \\
& \tilde{U} = U & \mbox{ if } & & U > b + \alpha_0 \\ 
\end{align*}
Then for any $x_1 \in [a,b]$, 
\begin{equation*}
\begin{aligned}
P(\tilde{U}  \leq x_1+\alpha|V=v) & = P(\tilde{U}  \leq x_1+\alpha, U \leq b + \alpha_0 |V=v) + P(\tilde{U}  \leq x_1+\alpha, U > b + \alpha_0|V=v) \\
&  = P(  U \leq x_1 + \alpha_0 |V=v) \\
P(\tilde{U}  \leq x_1|V=v) & = P(\tilde{U}  \leq x_1, U \leq b + \alpha_0 |V=v) + P(\tilde{U}  \leq x_1, U > b + \alpha_0|V=v) \\
&  = P(  U \leq b+\alpha_0, U \leq x_1 + \alpha_0 - \alpha |V=v ) + P(b+\alpha_0  < U \leq x_1 |V=v) \\
& =  P(  U \leq b+\alpha_0|V=v) + P(b+\alpha_0  < U \leq x_1 |V=v) \\
& =  P(U \leq x_1|V=v),
\end{aligned}
\end{equation*}
where the third equality holds because, since  $\alpha_0 < a-b$ and $\alpha < \alpha_0$, $b+\alpha_0 \leq x_1 + \alpha_0 - \alpha$ for $x_1 \in [a,b]$. Let $G_{U,V}$ and $G_{\tilde{U},V}$ be the joint distribution of $(U,V)$ and $(\tilde{U},V)$ respectively. Then the above calculation with \eqref{eq:without_P} imply that $(\alpha_0, G_{U,V})$ and $(\alpha, G_{\tilde{U},V})$ produce the identical pair $(P^{11}(x_1,x),P^{10}(x_1,x))$. In addition, the distribution of $V$ is unchanged so that $P(Y_2=1|X=x)$ is identified from data. Therefore, $(\alpha_0, G_{U,V})$ and $(\alpha, G_{\tilde{U},V})$ are observationally equivalent. 

\medskip
Similarly, when $\alpha_0 > b-a$, for any $\alpha > \alpha_0$, we can define
\begin{align*}
& \tilde{U} = U + \alpha- \alpha_0 & \mbox{ if } & & U > a + \alpha_0 \\
& \tilde{U} = U & \mbox{ if } & & U \leq a + \alpha_0 \\ 
\end{align*}
Then for any $x_1 \in [a,b]$, 
\begin{equation*}
\begin{aligned}
P(\tilde{U}  \leq x_1+\alpha|V=v) & = P(\tilde{U}  \leq x_1+\alpha, U \leq a + \alpha_0|V=v ) + P(\tilde{U}  \leq x_1+\alpha, U > a + \alpha_0|V=v) \\
&  = P(  U \leq a + \alpha_0 |V=v) +  P( a + \alpha_0 < U \leq x_1 + \alpha_0|V=v)\\
&  = P(U \leq x_1 + \alpha_0|V=v). \\
P(\tilde{U}  \leq x_1|V=v) & = P(\tilde{U}  \leq x_1, U \leq a + \alpha_0 |V=v) + P(\tilde{U}  \leq x_1, U > a + \alpha_0|V=v) \\
&  =  P(U \leq x_1|V=v),
\end{aligned}
\end{equation*}
where we use the facts that $x_1 \leq a + \alpha_0$ and $x_1 - a < \alpha$ for $x_1 \in [a,b]$. 
So again, $(\alpha_0, G_{U,V})$ and $(\alpha, G_{\tilde{U},V})$ are observationally equivalent.

\section{Proof of Theorem \ref{thm:partialid}}
\setcounter{equation}{0}
\label{sec:partialid}
The sign of $\alpha_0$ is identified by the data. In the following, we focus on deriving the results when $\alpha_0 > b-a$. By the proof of Theorem \ref{thm:id_suf}, we have already shown that all $\alpha > \alpha_0$ is in the identified set. Now we consider $\frac{b-a + \alpha_0}{2} \leq \alpha < \alpha_0$. 
\begin{align*}
& \tilde{U} = U + \alpha- \alpha_0 & \mbox{ if } & & U > a + \alpha \\
& \tilde{U} = U & \mbox{ if } & & U \leq a + \alpha \\ 
\end{align*}
Then for any $x_1 \in [a,b]$, 
\begin{equation*}
\begin{aligned}
P(\tilde{U}  \leq x_1+\alpha|V=v) & = P(\tilde{U}  \leq x_1+\alpha, U \leq a + \alpha |V=v) + P(\tilde{U}  \leq x_1+\alpha, U > a + \alpha|V=v) \\
&  = P(  U \leq a + \alpha |V=v) +  P( a + \alpha < U \leq x_1 + \alpha_0|V=v)\\
&  = P(U \leq x_1 + \alpha_0|V=v). \\
P(\tilde{U}  \leq x_1|V=v) & = P(\tilde{U}  \leq x_1, U \leq a + \alpha |V=v) + P(\tilde{U}  \leq x_1, U > a + \alpha|V=v) \\
&  =  P(U \leq x_1|V=v) + P(U \leq x_1 + \alpha_0 - \alpha, U > a + \alpha|V=v).\\
& =  P(U \leq x_1|V=v).
\end{aligned}
\end{equation*}
Here note that the last equality is because $x_1 + \alpha_0 - \alpha \leq b + \alpha_0 - \alpha \leq a + \alpha$ if $\alpha \geq \frac{b-a+\alpha_0}{2}$. Denote $\alpha^{(1)} = \frac{b-a+\alpha_0}{2}$. Then we have shown that there exists $U^{(1)}(\alpha)$ which only depends on $\alpha$ such that for any $x_1 \in [a,b]$, any $v$ and any $\alpha_0 >\alpha \geq \alpha^{(1)}$
\begin{equation*}
\begin{aligned}
& P(U^{(1)}(\alpha) \leq x_1 + \alpha|V=v) = P(U \leq x_1 + \alpha_0|V=v)\\
& P(U^{(1)}(\alpha) \leq x_1 |V=v) = P(U \leq x_1 |V=v).
\end{aligned}
\end{equation*}
In particular, there exists $U^{(1)}(\alpha^{(1)})$ such that 
\begin{equation*}
\begin{aligned}
& P(U^{(1)}(\alpha^{(1)}) \leq x_1 + \alpha^{(1)}|V=v) = P(U \leq x_1 + \alpha_0|V=v)\\
& P(U^{(1)}(\alpha^{(1)}) \leq x_1 |V=v) = P(U \leq x_1 |V=v).
\end{aligned}
\end{equation*}
Now repeating the above construction but replacing $U$ with $U^{(1)}$ and $\alpha_0$ with $\alpha^{(1)}$, we have for any $\alpha^{(1)} > \alpha \geq \alpha^{(2)} \equiv \frac{b-a + \alpha^{(1)}}{2}$, there exists $U^{(2)}(\alpha)$ such that for any $x_1 \in [a,b]$, any $v$ and any $\alpha^{(1)} >\alpha \geq \alpha^{(2)}$, 
\begin{equation*}
\begin{aligned}
& P(U^{(2)}(\alpha) \leq x_1 + \alpha^{(2)}|V=v) = P(U^{(1)}(\alpha^{(1)}) \leq x_1 + \alpha^{(1)}|V=v) = P(U \leq x_1 + \alpha_0|V=v)\\
& P(U^{(2)}(\alpha) \leq x_1 |V=v) = P(U^{(1)}(\alpha^{(1)}) \leq x_1 |V=v) = P(U \leq x_1 |V=v).
\end{aligned}
\end{equation*}
This concludes that any $\alpha$ such that $\alpha_0 > \alpha \geq \alpha^{(2)}$ is in the identified set. In general, by repeating the procedure k times, we have that any $\alpha$ such that 
$$\alpha_0 > \alpha \geq \alpha^{(k)} = (1-\frac{1}{2^k})(b-a) + \frac{\alpha_0}{2^k}$$
is in the identified set. For any $\alpha > b-a$, there exists some finite $k$ such that $\alpha > (1-\frac{1}{2^k})(b-a) + \frac{\alpha_0}{2^k}$. This concludes the result that $\alpha > b-a$ is in the identified set. 

\medskip
Finally, since if $\alpha > b-a$, $\partial_2 P^{11}(x_1,x) + \partial_2 P^{10}(\tilde{x}_1,x) > 0 $ for all pairs of $(x_1,x)$ and $(\tilde{x}_1,x)$ while, if $\alpha \leq b-a$, at least there exists one pair $(x_1,x)$ and $(\tilde{x}_1,x)$ such that $\partial_2 P^{11}(x_1,x) + \partial_2 P^{10}(\tilde{x}_1,x) \leq 0 $. This implies $\alpha \leq b-a$ is not in the identified set. Therefore, the sharp identified set when $\alpha_0 > b-a$ is $(b-a,\infty)$.  

\medskip
When $\alpha_0 < a-b$, a symmetric argument implies that the identified set is $(-\infty,a-b)$. 

\section{Estimation and Asymptotic Properties}\label{asymprop}
\setcounter{equation}{0}

Our identification result is constructive in the sense that it motivates an estimator for the parameters of interest which we describe in detail here.

As we did in Section \ref{sec:vys}, to simplify  exposition, in the following we focus exclusively on the parameters $\alpha_0,\gamma_0$.
%(\textbf{Yichong: The convergence rates for other parameters cannot exceed root-n (the convergence rate for $\alpha$). In addition, in order to estimate $(\beta_0,\lambda_0)$ and $\delta_0$, we need at least one element in $(Z_1,Z_3)$ and $Z$ to be %continuous, respectively. We assume $Z_2$ is continuous. Therefore, the later requirement is satisfied. But it is better if we can allow for $(Z_1,Z_3)$ to be discrete. In this case, $X_1 = Z_1'\lambda_0 + Z_3'\beta_0$ is discrete so that VY's estimator will not %work, as shown in the simulation. However, in this case, it is not obvious how can we estimate $(\beta_0,\lambda_0)$ is the first stage. I think we need to estimate $(\beta_0,\lambda_0)$ with $\alpha_0$ using our factor structure.})
Recall the choice probabilities $P^{ij}(x_1,x) = Prob(Y_1 = i,Y_2 = j|X_1 =x_1,X=x)$ and its second derivative  $\partial_2 P^{ij}(x_1,x)$, which can be estimated as we describe below. Another function needed for our identification result is the density function of the unobserved term $V$, denoted by $f_V(\cdot)$. This is also unknown,
but from the structure of our model can be recovered from the derivative with respect to the instrument $X$ of $E[Y_2|X]$, and hence is estimable from the data. Note that the proof of Theorem \ref{thm:factorid} shows that the sign of the index evaluated at two {\em different} regressor values, which we denote here by $(X_1,X)$ and $(\tilde X_1,\tilde X)$ is determined by the choice probabilities via
\[ 
\partial_2 P^{11}(X_1,X)/f_V(X) +\partial_2 P^{10}(\tilde X_1,\tilde X)/f_V(\tilde X) \geq 0  \ \ \ \iff X_1+\alpha-\gamma X-(\tilde X_1-\gamma \tilde X) \geq 0. \]
This motivates us to use the maximum rank correlation estimator proposed by \cite{MRC}.

Implementation requires further details to pay attention to. The unknown choice probabilities, their derivatives, and the density of $V$ will be estimated using nonparametric methods, and for this we adopt locally linear methods as they are particularly well suited for estimating derivatives of functions.

With functions and their derivatives estimated in the first stage of our procedure, the second stage plugs in these estimated values into an objective function to be optimized. Specifically,
letting $\hat \theta$ denote $(\hat \alpha,\hat \gamma)$, our estimator is of the form:
\begin{equation}
\label{rankestim}
\hat{\theta} = \arg\max_\theta Q_{n}(\theta), \quad Q_{n}(\theta) \equiv \sum_{i \neq j}\hat{g}_{i,j}(\theta)\end{equation}
in which
\begin{eqnarray*}
	\hat{g}_{i,j}(\theta) &= & [\mathbf{1}\{\partial_2 \hat{P}^{11}(X_{1,i},X_i)/\hat{f}_V(X_i) + \partial_2 \hat{P}^{10}(X_{1,j},X_j)/\hat{f}_V(X_j)\geq 0 \}\mathbf{1}\{\Phi(X_{1,i},X_i,X_{1,j},X_j;\theta) \geq 0\} \\
	& +& \mathbf{1}\{\partial_2 \hat{P}^{11}(X_{1,i},X_i)/\hat{f}_V(X_i) + \partial_2 \hat{P}^{10}(X_{1,j},X_j)/\hat{f}_V(X_j) <0\}\mathbf{1}\{\Phi(X_{1,i},X_i,X_{1,j},X_j;\theta) <0 \} ], 
\end{eqnarray*}
with 
\[\Phi(x_1,x,\tilde{x}_1,\tilde{x};\theta) = x_1 + \alpha - \gamma x - (\tilde{x}_1 - \gamma \tilde{x}).\]
We note that this estimator falls into the class of those which optimize a nonsmooth U-process involving components estimated nonparametrically in a preliminary stage.\footnote{An alternative
	estimation procedure could be based on the exact relationship in \eqref{eq:match}. Note the equality on the left-hand side of \eqref{eq:match} is a function of the data alone and not the unknown parameters.
	%Furthermore, as said, while these functions, choice probability, density functions are unknown they can be consistently estimated from the data in a preliminary stage.
	The right-hand side equality can then be regarded as a moment condition to estimate the unknown parameters. We describe this estimator and derive its asymptotic properties in the Online Supplement to the paper. While the two estimation approaches will have similar asymptotic properties (root-$n$ consistent, asymptotically normal), we prefer the rank estimator in (\ref{rankestim}) which involves fewer tuning parameters. Furthermore rank type estimators in general are more robust to certain types of misspecification, as pointed out in \cite{khantamerjbes}.}
Examples of other estimators in this class can be found in \cite{khan2001}, \cite{abhausmankhan}, \cite{jochmansej}, \cite{chenkhantang},
and our approach to deriving the limiting distribution theory of our estimator will follow along the steps used in those papers.
%Our proof strategy will be based on deriving a quadratic approximation for the objective function $Q_{n}(\theta)$, in a way analogous to the method introduced
%in \cite{sherman-et}. Following \cite{sherman-et}, we will derive the asymptotic properties of $\hat \theta$ in three stages.  We will first establish its consistency, then derive an intermediate rate ($4^{th}$ root consistency),
%followed by establishing root-$n$ consistency and asymptotic normality of the estimator. 
Our limiting distribution theory for this estimator is  based on the following regularity conditions:
\begin{description}
	\item[RK1] $\theta_0$ lies in the interior of $\Theta$, a compact subset of $R^2$.
	\item[RK2]  The index $X$ is continuously distributed with support on the real line, and has a density function which is twice continuously differentiable.
	\item [RK3] (Order of smoothness of probability functions and regressor density functions) The functions $P^{i,j}(\cdot)$
	and $f_{X_1,X}(\cdot.\cdot)$ (the density function of the random vector $(X_1,X)$) are continuously differentiable
	of order $p_2$.
	\item [ RK4] (First stage kernel function conditions) $K(\cdot)$, used to estimate the choice probabilities and their derivatives is an even function, integrating to 1 and is of order  $ p_2$.
	\item [ RK5] (Rate condition on first stage bandwidth sequence) The first stage  bandwidth sequence $H_n$ used in the nonparametric estimator of the choice probability functions and their derivatives
	satisfies $\sqrt{n}H_n^{p_2-1}\rightarrow 0$ and $n^{-1/4}H_n^{-1}\rightarrow 0$.
\end{description}

The smoothness condition in Assumption RK4 and Assumption RK5 is due to the fact that we need to nonparametrically estimate $\partial_2P^{ij}(X_1,X)$ with sufficiently faster convergence rate. This will require a stronger smoothness condition than that required for standard nonparametric estimation. Assumption RK5 ensures that the bias of the first stage estimator of the derivative function converges at the parametric rate
and the RMSE of this estimator (with two regressors) is fourth-root consistent, so results for two step estimation in  \cite{newey_mcfadden} can be applied. 

Based on these conditions, we have the following theorem, whose proof is in Section \ref{sec:asymp} of the Supplementary Appendix which characterizes the rate of convergence and asymptotic distribution of the proposed estimator:
\begin{theorem}
	\label{thm:rank}
	Under Assumptions {\bf RK1-RK5},
	\begin{equation}
	\sqrt{n}(\hat \theta-\theta_0) \Rightarrow N ( 0 , V^{-1} \Delta V^{-1} ) 
	\end{equation}
	where the forms of the Hessian term $V$ and outer score term $\Delta$ are described in detail in Section \ref{sec:asymp} of the Supplementary Appendix.
	%\begin{equation}
	%V=\nabla_{\theta\theta} {\cal G}(\theta_0)
	%\end{equation}
	%and 
	%\begin{equation}
	%\Delta= E\left[ (\psi_{1rnki}+\psi_{2rnki})(\psi_{1rnki}+\psi_{2rnki})' \right]
	%\end{equation}
\end{theorem}

\section{Proof of Theorem \ref{thm:rank}}
\setcounter{equation}{0}
\label{sec:asymp}
Recall we defined our two step rank estimator as follows:
Letting $\hat \theta$ denote $(\hat \alpha,\hat \gamma)$, our estimator is of the form:

\[\hat{\theta} = \arg\max_\theta \hat Q_{n}(\theta) \equiv \sum_{i \neq j}\hat{g}_{i,j}(\theta)\]
in which
\begin{eqnarray*}
	\hat{g}_{i,j}(\theta) &= & [\mathbf{1}\{\partial_2 \hat{P}^{11}(X_{1,i},X_i)/\hat{f}_V(X_i) + \partial_2 \hat{P}^{10}(X_{1,j},X_j)/\hat{f}_V(X_j)\geq 0 \}\mathbf{1}\{\Phi(X_{1,i},X_i,X_{1,j},X_j;\theta) \geq 0\} \\
	& +& \mathbf{1}\{\partial_2 \hat{P}^{11}(X_{1,i},X_i)/\hat{f}_V(X_i) + \partial_2 \hat{P}^{10}(X_{1,j},X_j)/\hat{f}_V(X_j) <0\}\mathbf{1}\{\Phi(X_{1,i},X_i,X_{1,j},X_j;\theta) <0 \} ], 
\end{eqnarray*}
with 
\[\Phi(x_1,x,\tilde{x}_1,\tilde{x};\theta) = x_1 + \alpha - \gamma x - (\tilde{x}_1 - \gamma \tilde{x})\]

We first show consistency of the rank estimator. To do so
we first define the  objective function $Q^{if}_{n,2}(\theta)$, defined as 
\[ Q^{if}_{n,2}(\theta) \equiv \sum_{i \neq j}{g}_{i,j}(\theta) \]
where
\begin{eqnarray*}
	{g}_{i,j}(\theta) &= & [\mathbf{1}\{\partial_2 {P}^{11}(X_{1,i},X_i)/{f}_V(X_i) + \partial_2 {P}^{10}(X_{1,j},X_j)/{f}_V(X_j)\geq 0 \}\mathbf{1}\{\Phi(X_{1,i},X_i,X_{1,j},X_j;\theta) \geq 0\} \\
	& +& \mathbf{1}\{\partial_2 {P}^{11}(X_{1,i},X_i)/{f}_V(X_i) + \partial_2 {P}^{10}(X_{1,j},X_j)/{f}_V(X_j) <0\}\mathbf{1}\{\Phi(X_{1,i},X_i,X_{1,j},X_j;\theta) <0 \} ], 
\end{eqnarray*}
Since $g_{i,j}$ is  bounded by 1 $\forall i,j$, and our random sampling assumption, we have  for each $\theta$,
\[  Q^{if}_{n,2}(\theta)\stackrel{p}{\rightarrow} E[g_{i,j}(\theta)]\equiv \Gamma_0(\theta) \]
Furthermore, by Assumptions RK2, RK3 we can extend this result to converging uniformly over $\theta\in \Theta$ (see, e.g. \cite{sherman-annals}, \cite{sherman}.)
$\Gamma_0(\theta)$ is continuous in $\theta$ by Assumptions RK2,RK3, and uniquely maximized at $\theta=\theta_0$ by our identification result in Theorem \ref{thm:factorid}. 
Along with Assumption RK1, the infeasible estimator, defined as the maximizer of  $ Q^{if}_{n,2}(\theta)$ converges in probability to $\theta_0$
by, for example Theorem 2.1 in  \cite{newey_mcfadden}.
To show consistency of the feasible estimator, where we first estimate the choice probability functions and their derivatives nonparametrically, we only now need to show the two objective functions converged to each other uniformly in $\theta\in \Theta$.
Consistency of the first stage estimators follows from Assumptions {\bf RK3-RK5}, see for example \cite{liparmeter}.
However, this does not immediately imply convergence of the difference in feasible and infeasible objective functions since the nonparametric estimators are inside indicator functions so the continuous mapping theorem does immediately not apply. Nonetheless the desired result can still be attained in one of two ways. One would be to replace indicator functions with smooth distribution functions in a fashion analogous to \cite{horowitz-sms}. This would have the disadvantage of introducing tuning parameters, but another approach would be to replace the indicator functions with their conditional expectations, and note that the conditional expectations are smooth functions using Assumption {\bf RK2, RK3}. To see why, let $\hat m(x_i)$ be a nonparametric estimator of a function $m(x_i)$, which is assumed to be smooth. We evaluate the plim of \[I[\hat m(x_i)>0]-I[m(x_i)>0]=I[\hat m(x_i)>0,m(x_i)<0]-I[\hat m(x_i)<0,m(x_i)>0]\]
we show that the first term converges in probability to 0 as identical arguments can be used for the second term.
Let $\varepsilon>0$ be given; $P(I[\hat m(x_i)>0,m(x_i)<0]>\varepsilon)\leq E[I[\hat m(x_i)>0,m(x_i)<0]/\varepsilon$ by Markov's inequality.
But the expectation in the numerator on the right hand side is
\[ P(\hat m(x_i)>0, m(x_i)<0)=P(\hat m(x_i)>0, m(x_i)\leq -\delta_n)+P(\hat m(x_i)>0, m(x_i)\in (-\delta_n,0)) \]
where $\delta_n$ is a sequence of positive  numbers converging to 0, at a slow rate, e.g.$(\log n^{-1})$.
The first term on the right hand side is bounded above by
\[ P(|\hat m(x_i)-m(x_i) |>\delta_n)\leq P(\|\hat m(\cdot)-m(\cdot)\|>\delta_n) \]
where the notation $\| \hat m(\cdot)-m(\cdot)\|$ above denotes the sup norm over $x_i$.
The right hand side  probability above will be sufficiently small for $n$ large enough by the rate of convergence of the nonparametric estimator.
The second term, $P(\hat m(x_i)>0, m(x_i)\in (-\delta_n,0))$, is bounded above by $P(m(x_i)\in (-\delta_n,0))$ which by the smoothness of $m(x_i)$  converges to 0, and hence can be made arbitrarily small.
$\hfill \square$ 

%We next expand the function $\Gamma_0(\theta)$ around $\theta=\theta_0$ which is permitted by Assumption RS. Note the linear term in a second order expansion of $\Gamma_0(\theta)$
%is 0 at $\theta=\theta_0$ by our identification result in Theorem \ref{identificationrank}. Thus we have 
%\begin{equation} \label{expandlimobj} \Gamma_0(\theta)=\Gamma_0(\theta_0)+\frac{1}{2} (\theta-\theta_0)'\nabla_2 \Gamma_0(\theta_0) (\theta-\theta_0)+ o(||\theta-\theta_0||^2) \end{equation}
%where $\nabla_2$ denotes the second derivative operator and $||\cdot||$ denotes the Euclidean norm.

To derive the rate of convergence and limiting distribution theory for the feasible estimator where we first estimate choice probability functions
and their derivatives nonparametrically, we  expand the nonparametric estimators around true functions that are inside the indicator function in $Q_{n2}$.
Then we can  follow the approach in \cite{sherman-et}. Having already
established consistency of the estimator, we will first establish
root-$n$ consistency and then  asymptotic normality. For root-$n$
consistency we will apply Theorem 1 of \cite{sherman-et} and so here we change notation to deliberately stay as close as possible to his. We will
actually apply this theorem twice, first establishing a slower
than root-$n$ consistency result and then root-$n$ consistency.
Keeping our notation deliberately as close as possible  to Sherman(1994b),
here replacing our second stage rank objective function $\hat{Q}_{2,n}(\theta)$
with
$\hat{\cal G}_n(\theta)$, our infeasible objective function $ Q^{if}_{n,2}(\theta)$ with  ${\cal G}_n(\theta)$,  and denoting our limiting objective function,
previously denoted by ${\Gamma_0}(\theta)$,
by ${\cal G}(\theta)$. We have the following theorem:
\begin{theorem}\label{sherman1}
	(From Theorem 1 in \cite{sherman-et}). 
	
	\noindent
	If $\delta_n$ and $\epsilon_n$ are sequences of positive numbers converging to 0, and 
	\begin{enumerate}
		\item $\hat{\theta}-\theta_0=o_p(\delta_n)$ \item There exists a
		neighborhood of $\theta_0$ and a constant $\kappa>0$ such that
		${\cal G}(\theta)-{\cal G}(\theta_0)\geq \kappa \|\theta-\theta_0\|^2$
		for all $\theta$ in this neighborhood. \item Uniformly over
		$O_p(\delta_n)$ neighborhoods of $\theta_0$
		\[ \hat{\cal G}_n(\theta)={\cal G}(\theta)+ O_p(\|\theta-\theta_0\|/\sqrt{n})
		+o_p(\|\theta-\theta_0\|^2)+O_p(\varepsilon_n) \]
	\end{enumerate}
	then
	$\hat{\theta}-\theta_0=O_p(\max(\varepsilon^{1/2},n^{-1/2}))$. 
\end{theorem}
Once we use this theorem to establish the rate of convergence of our rank estimator, we can attain limiting distribution theory, which will follow from the following theorem:
\begin{theorem}\label{sherman2}
	(From Theorem 2 in \cite{sherman-et}). 
	Suppose $\hat \theta$ is $\sqrt{n}$-consistent for $\theta_0$, an interior point of $\Theta.$ Suppose also that uniformly over $O_p(n^{-1/2})$ neighborhoods of $\theta_0$,
	\begin{equation}
	\hat {\cal G}_n(\theta) = \frac{1}{2} (\theta-\theta_0)' V(\theta-\theta_0) + \frac{1}{\sqrt{n}}(\theta-\theta_0)' W_n + o_p(1/n)
	\end{equation}
	where $V$ is a negative definite matrix, and $W_n$ converges in distribution to a $N(0,\Delta)$ random vector. Then
	\begin{equation}
	\sqrt{n}(\hat \theta-\theta_0) \Rightarrow N ( 0 , V^{-1} \Delta V^{-1} ) 
	\end{equation}
\end{theorem}
We first turn attention to applying Theorem \ref{sherman1} to derive the rate of convergence of our estimator.
Having already established consistency of our rank estimator, we turn attention to the second condition in Theorem \ref{sherman1}.
To
show the second condition, we will first derive an expansion for
${\cal G}(\theta)$ around ${\cal G}(\theta_0)$. We denote that even
though ${\cal G}_n(\theta)$ is not differentiable in $\theta$,
${\cal G}(\theta)$ is sufficiently smooth for Taylor expansions to
apply as the expectation operator is a smoothing operator and the
smoothness conditions in Assumptions {\bf RK2}, {\bf RK3}.
Taking a second order expansion of ${\cal G}(\theta)$ around ${\cal
	G}(\theta_0)$, we obtain
\begin{equation}
{\cal G}(\theta)={\cal G}(\theta_0)+\nabla_\beta {\cal
	G}(\theta_0)'(\theta-\theta_0) +\frac{1}{2}
(\theta-\theta_0)'\nabla_{\theta\theta} {\cal
	G}(\theta^\ast)(\theta-\theta_0)
\end{equation}
where $\nabla_\theta$ and $\nabla_{\theta\theta}$ denote first and
second derivative operators and $\theta^\ast$ denotes an
intermediate value. We note that the first two terms of the right
hand side of the above equation are 0, the first by how we defined
the objective function, and the second by our identification
result in Theorem \ref{thm:factorid}. Define
\begin{equation}
V\equiv \nabla_{\theta\theta} {\cal G}(\theta_0)
\end{equation}
and $V$ is positive definite by Assumption {\bf A3}, so we have
\begin{equation}
(\theta-\theta_0)'\nabla_{\theta\theta} {\cal
	G}(\theta_0)(\theta-\theta_0) >0
\end{equation}
$\nabla_{\theta\theta}{\cal G}(\theta)$ is also continuous at
$\theta=\theta_0$ by Assumptions {\bf RK2} and {\bf RK3}, so there
exists a neighborhood of $\theta_0$ such that for all $\theta$ in
this neighborhood, we have
\begin{equation}
(\theta-\theta_0)'\nabla_{\theta\theta} {\cal G}(\theta)(\theta-\theta_0)
>0
\end{equation}
which  suffices for the second condition to hold.

To show the third condition in Theorem \ref{sherman1}, we next establish the form of the remainder term when we replace nonparametric estimators with the true functions they are estimating.

Specifically we wish to evaluate the difference between
{\scriptsize \begin{eqnarray} \label{nonpest1}
	&\ & [\mathbf{1}\{\partial_2 \hat{P}^{11}(X_{1,i},X_i)/\hat{f}_V(X_i) + \partial_2 \hat{P}^{10}(X_{1,j},X_j)/\hat{f}_V(X_j)\geq 0 \}\mathbf{1}\{\Phi(X_{1,i},X_i,X_{1,j},X_j;\theta) \geq 0\} \\
	& +& \mathbf{1}\{\partial_2 \hat{P}^{11}(X_{1,i},X_i)/\hat{f}_V(X_i) + \partial_2 \hat{P}^{10}(X_{1,j},X_j)/\hat{f}_V(X_j) <0\}\mathbf{1}\{\Phi(X_{1,i},X_i,X_{1,j},X_j;\theta) <0 \} 
	\end{eqnarray}}
and
{\scriptsize \begin{eqnarray} \label{truefun1}
	&\  & [\mathbf{1}\{\partial_2 {P}^{11}(X_{1,i},X_i)/{f}_V(X_i) + \partial_2 {P}^{10}(X_{1,j},X_j)/{f}_V(X_j)\geq 0 \}\mathbf{1}\{\Phi(X_{1,i},X_i,X_{1,j},X_j;\theta) \geq 0\} \\
	& +& \mathbf{1}\{\partial_2 {P}^{11}(X_{1,i},X_i)/{f}_V(X_i) + \partial_2 {P}^{10}(X_{1,j},X_j)/{f}_V(X_j) <0\}\mathbf{1}\{\Phi(X_{1,i},X_i,X_{1,j},X_j;\theta) <0 \}  
	\end{eqnarray}}
To establish a representation for this difference, we first simplify notation we write the expressions as:
\begin{eqnarray}
&\ &I[\hat m_1({\bf x}_i)+\hat m_2({\bf x}_j)\geq 0]I[\Delta {\bf x}_{ij}'\theta\geq 0] \\
&+& I[\hat m_1({\bf x}_i)+\hat m_2({\bf x}_j)< 0]I[\Delta {\bf x}_{ij}'\theta< 0] 
\end{eqnarray}
and
\begin{eqnarray}
&\ &I[ m_1({\bf x}_i)+ m_2({\bf x}_j)\geq 0]I[\Delta {\bf x}_{ij}'\theta\geq 0] \\
&+& I[ m_1({\bf x}_i)+ m_2({\bf x}_j)< 0]I[\Delta {\bf x}_{ij}'\theta< 0] 
\end{eqnarray}
respectively, where here ${\bf x}_i$ denotes the separate components of $x_{1i},x_i$, and analogous for ${\bf x}_j$. We first explore
\[ (I[\hat m_1({\bf x}_i)+\hat m_2({\bf x}_j)\geq 0]-I[ m_1({\bf x}_i)+ m_2({\bf x}_j)\geq 0])I[\Delta {\bf x}_{ij}'\theta\geq 0]  \]
for each $i,j$ inside the double summation:
\begin{equation}\label{indicatediff}
\usum (I[\hat m_1({\bf x}_i)+\hat m_2({\bf x}_j)\geq 0]-I[ m_1({\bf x}_i)+ m_2({\bf x}_j)\geq 0])I[\Delta {\bf x}_{ij}'\theta\geq 0] 
\end{equation}

An immediate technical difficulty that arises with the above term is the presence of a nonparametric estimator inside the indicator function above.
A simple approach to deal with this would be to replace the indicator function  with a smoothed indicator function in a fashion analogous
to \cite{horowitz-sms}, under appropriate conditions on the kernel function and smoothing parameter. Such an approach is not necessary
as long as the nonparametric estimator $\hat m_1(x_i)$ is asymptotically normal, and asymptotically centered at $m_1(x_i)$, which will be the case with our proposed kernel estimator of the probability function and its derivative. In either approach (smoothed indicator or not) we can show that
(\ref{indicatediff}) can be represented as:
\begin{equation}\label{indicatediffb}
\usum \phi(0)f_{m_{ij}}(0) \left( (\hat m_1({\bf x}_i)-m_1({\bf x}_i))+(\hat m_2({\bf x}_j)-m_2({\bf x}_j))\right)I[\Delta {\bf x}_{ij}'\theta\geq 0] 
+o_p(n^{-1})
\end{equation}
where $\phi(0)$ denotes the standard normal pdf evaluated at 0, $f_{m_{ij}}(0)$ denotes the density function of
$m_1({\bf x}_i)+ m_2({\bf x}_j)$ evaluated at 0,     and  the $o_p(n^{-1})$ term is uniform in $\theta$ lying in $o_p(1)$ neighborhoods of $\theta_0$.
Therefore, uniformly for $\theta$ in  an $o_p(1)$ neighborhood of $\theta_0$, this remainder term converges to 0 at the rate of convergence
of the first stage nonparametric estimator, which under Assumptions RK3, RK4, RK5, is $o_p(n^{-1/4})$.
Thus by repeated application of Theorem \ref{sherman1}, we can conclude that the estimator is root-$n$ consistent.
To show that the estimator is also asymptotically normal, 
we will first derive a linear representation for the term:
\begin{equation}\label{khan2001}
\usum \phi(0)f_{m_{ij}}(0)  (\hat m_1({\bf x}_i)-m_1({\bf x}_i))I[\Delta {\bf x}_{ij}'\theta\geq 0] 
\end{equation}
As this term is linear in the nonparametric estimator $\hat m_1(x_i)$, the desired linear 
representation follows  from arguments used in \cite{khan2001}. One slight difference here compared to \cite{khan2001} is that here our nonparametric 
estimators and estimands are each ratios of derivatives. Nonetheless, after linearizing these ratios
as done in, e.g. \cite{newey_mcfadden}.  Specifically, we have that \ref{khan2001} can be expressed as:
\begin{equation}\label{khan2001a}
\usum \phi(0)f_{m_{ij}}(0) \frac{1}{m_{1den}({\bf x}_i)} (\hat m_{1num}({\bf x}_i)-m_{1num}({\bf x}_i))I[\Delta {\bf x}_{ij}'\theta\geq 0]  
\end{equation}
\begin{equation}\label{khan2001b}
-\usum \phi(0)f_{m_{ij}}(0) \frac{m_{1num}({\bf x}_i)}{m_{1den}({\bf x}_i)^2} (\hat m_{1den}({\bf x}_i)-m_{1den}({\bf x}_i))I[\Delta {\bf x}_{ij}'\theta\geq 0]  
\end{equation}
where $\hat m_{1num}({\bf x}_i)$ denotes the numerator $ \{\partial_2 \hat{P}^{11}(X_{1,i},X_i)\}$,
the estimator of $ m_{1num}({\bf x}_i)$ which denotes $\{\partial_2 {P}^{11}(X_{1,i},X_i)\}$,
and $\hat m_{1den}({\bf x}_i)$ denotes the denominator $\hat{f}_V(X_i)$, the estimator of  $m_{1den}({\bf x}_i)$ which denotes
${f}_V(X_i)$.

Plugging in the definitions of the kernel estimators of $\hat m_{1num}({\bf x}_i)$, and $\hat m_{1den}({\bf x}_i)$,
results in a third order process. Using arguments in \cite{khan2001} and \cite{powell_stock_stoker} we can express the third order $U$ process as a second order $U$ process plus 
an asymptotically negligible remainder term. This is of the form:
\begin{equation}\label{khan2001c}
\avg \phi(0) \frac{\ell(x_i)}{m_{1den}({\bf x}_i)} ( y_{1i}-m_{1num}({\bf x}_i))E\left[I[f_{m_{ij}}(0)\Delta {\bf x}_{ij}'\theta\geq 0]|x_i\right] 
\end{equation}
where $\ell(x_i)\equiv \frac{-f_X'(x_i)}{f_X(x_i)}$.
We note that the function $E\left[f_{m_{ij}}(0)I[\Delta {\bf x}_{ij}'\theta\geq 0]|x_i\right] $, which we denote here by ${\cal H}(x_i,\theta)$ is a smooth function in $\theta$. We will use this feature to expand
${\cal H}(x_i,\theta)$  around  ${\cal H}(x_i,\theta_0)$.
Analogous arguments can be used to attain a linear representation of (\ref{khan2001b}), which is of the form:
\begin{equation}\label{khan2001d}
\avg \phi(0) \frac{\ell_2(x_{1i})m_{1num}({\bf x}_i)}{m_{1den}({\bf x}_i)^2} ( y_{2i}-m_{1den}({\bf x}_i))E\left[I[f_{m_{ij}}(0)\Delta {\bf x}_{ij}'\theta\geq 0]|x_i\right] 
\end{equation}
where $\ell_2(x_{1i})\equiv \frac{-f_{X_1}'(x_{1i})}{f_X(x_{1i})}$.
Grouping (\ref{khan2001c}) and (\ref{khan2001d}) we have
\begin{equation}\label{influencernk1}
\avg \phi(0)   \frac{1}{m_{1den}({\bf x}_i)} \left\{\ell(x_i)( y_{1i}-m_{1num}({\bf x}_i))-\frac{m_{1num}({\bf x}_i)}{m_{1den}({\bf x}_i)} \ell_2(x_{1i})( y_{2i}-m_{1den}({\bf x}_i))\right\}{\cal H}(x_i,\theta)
\end{equation}
Note that by Assumptions {\bf RK2}, {\bf RK3}, $ {\cal H}(x_i,\theta)$ is smooth in $\theta$ implying the expansion
\[  {\cal H}(x_i,\theta)={\cal H}(x_i,\theta_0) + \nabla_\theta {\cal H}(x_i,\theta_0)'(\theta-\theta_0) \]
Thus we can express (\ref{influencernk1}) as the 
which we note is a mean 0 sum
\begin{equation} \label{infrnk1}
\avg \psi_{1rnki} (\theta-\theta_0)
\end{equation}
where {\small
	\begin{equation}
	\psi_{1rnki}=\phi(0)   \frac{1}{m_{1den}({\bf x}_i)} \left\{\ell(x_i)( y_{1i}-m_{1num}({\bf x}_i))-\frac{m_{1num}({\bf x}_i)}{m_{1den}({\bf x}_i)} \ell_2(x_{1i})( y_{2i}-m_{1den}({\bf x}_i))\right\}\nabla_\theta{\cal H}(x_i,\theta_0)
	\end{equation}}
We can use identical arguments to attain a linear representation for the $U-$ process:
\begin{equation}\label{indicatediffc}
\usum \phi(0)f_{m_{ij}}(0) \left( \hat m_2({\bf x}_j)-m_2({\bf x}_j)\right)I[\Delta {\bf x}_{ij}'\theta\geq 0] 
\end{equation}
where $\hat m_2({\bf x}_j)$ is also a ratio of nonparametric estimators
where here the numerator is  $\hat m_{2n}({\bf x}_j)$ denoting  $ \{\partial_2 \hat{P}^{10}(X_{1,j},X_j)\}$,
the estimator of $ m_{2n}({\bf x}_2)$ which denotes $\{\partial_2 {P}^{10}(X_{1,j},X_j)\}$,
and $\hat m_{2d}({\bf x}_j)$ denotes the denominator $\hat{f}_V(X_j)$, the estimator of  $m_{1den}({\bf x}_j)$ which denotes
${f}_V(X_j)$.

and by using identical arguments it too can be represented as a mean 0 sum denoted here by
\begin{equation} \label{infrnk2}
\avg \psi_{2rnki}
\end{equation}
where $\psi_{2rnki}$ is defined as:

Finally after grouping the two terms and expanding ${\cal H}(x_i,\theta)$ around ${\cal H}(x_i,\theta_0)$ we get that (\ref{indicatediffb})
can be represented as:
\begin{equation} \label{influencesheman}
\avg (\psi_{1rnki}+\psi_{2rnki})'(\theta-\theta_0) +o_p(n^{-1})
\end{equation}

Combining our results, from Theorem \ref{sherman2}, we have that
\begin{equation}
\sqrt{n}(\hat \theta-\theta_0) \Rightarrow N ( 0 , V^{-1} \Delta V^{-1} ) 
\end{equation}
where 
\begin{equation}
V=\nabla_{\theta\theta} {\cal G}(\theta_0)
\end{equation}
and 
\begin{equation}
\Delta= E\left[ (\psi_{1rnki}+\psi_{2rnki})(\psi_{1rnki}+\psi_{2rnki})' \right]
\end{equation}

\section{Model with Two Idiosyncratic Shocks}
\setcounter{equation}{0}
\label{App:bounded_factor}
In this section, we focus on the identification of $(\alpha_0,\gamma_0)$ in the  ``condensed" model that $X_1 = Z_1'\lambda_0 + Z_3'\beta_0$ is observed and 
\begin{equation}
\begin{aligned}
& Y_1 = \mathbf{1}\{X_1 + \alpha_0 Y_2 - U \geq 0 \} \\
& Y_2 = \mathbf{1}\{X - V \geq 0 \}.
\label{eq:2Fdis'}
\end{aligned}
\end{equation}
with the understanding that $(\lambda_0,\beta_0)$ can be identified jointly with $\alpha_0$ and $\gamma_0$, as shown in Theorems \ref{thm:factorid} and \ref{thm:aux}.  We further impose $U = \gamma_0 W + \eta_1$, $V = W + \eta_2$, and $(W, \eta_1,\eta_2)$ are mutually independent. First we consider the case $\gamma_0 = 1$ and $X_1$ is binary, because even in this context, for the baseline case with one idiosyncratic shock, we can identify $\alpha_0$. 
But identification of $\alpha_0$ becomes more difficult in this model without the help of repeated measurements, as established in the following theorem.
\begin{theorem}
	\label{thm:2Fids}
	Suppose \eqref{eq:2Fdis'} holds, $\gamma_0$ is known to be one, $X_1$ is binary, and $W$ has a bounded support $[-b,-a]$ such that $0.5>b-a$ and $1-(b-a) > \alpha_0 > b-a$, then $\alpha_0$ is {\bf not} point identified.
\end{theorem}
This nonidentification result motivates imposing additional structure on $W$, and we consider the following model
\begin{description}
	\item[C1] $U = \gamma_0 W + \eta_1 \quad \text{and} \quad V = \sigma_0 W + \eta_2$.
	\item[C2] $W$ is standard normally distributed. 
	\item[C3] $W$, $\eta_1$ and $\eta_2$ are mutually independent. 
	\item[C4] $X$ has full support. 
	\item[C5] Denote the density of $\eta_2$ as $f_{\eta_2}$, then $f_{\eta_2}$ does not have a Gaussian component in the sense that
	\begin{align*}
	f_{\eta_2} \in \mathcal{G} = \{g~\text{is a density on }\Re~\text{s.t.}: g = g'*\phi_\sigma~\text{for some density $g'$ implies that }\sigma=0\},
	\end{align*} 
	where $\phi_\sigma$ is the density for a normal distribution with zero mean and $\sigma^2$ variance.
\end{description}

Assumption \textbf{C5} effectively assumes that the distribution of $\eta_2$ has tail properties different from those of a normal distribution. 
This type of assumption is made in the deconvolution literature as it is necessary for identification of the target density when the error distribution is not completely known-
see, e.g., \cite{butucea05}.\footnote{In fact, based on the results  in \cite{butucea05}, $W$ can belong to a more general class of known distributions.
	Furthermore, we note that if  $\sigma_0$ is known, then Assumption \textbf{C5} is not necessary.} The importance of non-normality in factor models goes back to \cite{Geary42} and \cite{reiersol}, who have shown that factor loadings are identified in a linear measurement error model if the factor is not Gaussian. In our case, note $V = \sigma_0 W+\eta_2$ where $W$ is standard normal and the density of $V$ is identified from data. Here we want to identify $\sigma_0$ and the density of $\eta_2$. If $\eta_2$ has a Gaussian component, then 
\begin{align*}
\eta_2 = \eta_2' + \tilde{\sigma} \tilde{W},
\end{align*}
where $\tilde{W}$ is a standard normal random variable that is independent of $\eta_2'$ and $W$ and $\tilde{\sigma} >0$. It implies 
\begin{align*}
V = (\sigma_0 W + \tilde{\sigma} \tilde{W}) + \eta_2',
\end{align*}
where $\eta_2'$ does not have a Gaussian component. In addition, note that $(\sigma_0 W + \tilde{\sigma} \tilde{W}) = \sqrt{\sigma_0^2 + \tilde{\sigma}^2} G$, for some standard normal random variable $G$. Therefore, without Assumption \textbf{C5}, $\sigma_0$ is not identified.

\begin{theorem}
	\label{thm:twofactor2}
	If Assumptions \textbf{C1}--\textbf{C5} hold, then $\sigma_0$, $\gamma_0$ and $\alpha_0$ are identified. 
\end{theorem}
Note that this identification result does not require any variation from $X_1$, which is in spirit close to the one-factor model in our paper and is different from the identification result in \cite{vytlacilyildiz}. 
We also  note that this result does not contradict the counterexample in the paper. In the counterexample, we only assume that we know the support of $W$ is bounded. Here we assume that the full density of $W$, and thus, the support of $W$ is known.

%\section{Partial Identification}
%\label{sec:partialid}
%For illustration purpose, we focus on the ``condensed" model:
%\begin{equation*}
%\begin{aligned}
%& Y_1 = \mathbf{1}\{X_1 + \alpha_0 Y_2 - U \geq 0 \} \\
%& Y_2 = \mathbf{1}\{X - V \geq 0 \}.
%\end{aligned}
%\end{equation*}
%
%\begin{theorem}	
%	Assumption \ref{ass:indep} holds. When $|\alpha_0|> b-a$, the sharp identified set for $\alpha_0$ is 
%	$$\mathcal{A}^* = \{\alpha: \alpha > b-a \mbox{ if } \alpha_0 > 0 \mbox{ and } \alpha < a-b \mbox{ if } \alpha_0 < 0  \}.$$ 
%	\label{thm:partialid}
%\end{theorem}

\section{Proof of Theorem \ref{thm:2Fids}}
\setcounter{equation}{0}
\label{sec:2Fids}
Our first result for this model illustrates how identification can become more difficult. In our first result for this model, we show when $-W$ has a bounded support, say $[a,b]$, then $\alpha_0$ is not identified if $\alpha_0 > b-a$. 
To establish this, consider an impostor $\alpha$ such that $\alpha < \alpha_0$. In addition, we consider the case where $\alpha_0 - \alpha + b < \alpha_0 + a$ and $\alpha + b < a+1$. Such $\alpha$ exists because of the fact that $1-(b-a) > \alpha_0 > b-a.$ Let $\Delta = \alpha_0 - \alpha$ and $(\tilde{W},\tilde{\eta}_1,\tilde{\eta}_2)$ be mutually independent such that $\tilde{W}$ is distributed as  $W - \Delta$, 
$\tilde{\eta}_2$ is distributed as  $\eta_2 - \Delta$, and 
$$
F_{\tilde{\eta}_1}(e) =
\begin{cases}
F_{\eta_1}(e) & \text{ on } e \leq a,\\
F_{\eta_1}(a)  & \text{ on } \eta_1 \in (a,a+\Delta],\\
F_{\eta_1}(e - \Delta) & \text{ on } e \in (a+\Delta,b+\Delta],\\
\frac{\alpha_0 + a -e}{\alpha_0 + a -b - \Delta}F_{\eta_1}(b) + \frac{e - b - \Delta}{\alpha_0 + a -b - \Delta}F_{\eta_1}(\alpha_0 + a) & \text{ on } e \in (b+\Delta,\alpha_0+a], \\
F_{\eta_1}(e) & \text{ on } e \in (\alpha_0+a,\alpha_0+b),\\
F_{\eta_1}(\alpha_0+ b) + \frac{e - \alpha_0 - b}{a+1+\Delta - \alpha_0 - b}(F_{\eta_1}(a+1) - F_{\eta_1}(\alpha_0 + b)) & \text{ on } e \in (\alpha_0+b,a+1+\Delta],\\
F_{\eta_1}(e - \Delta) & \text{ on } e \in (a+\Delta+1,b+\Delta+1],\\
F_{\eta_1}(b+1)+ \frac{e - (b+\Delta + 1)}{a+ \alpha_0 - b-\Delta}(F_{\eta_1}(a + \alpha_0 + 1) - F_{\eta_1}(b+1)) &  \text{ on } e \in  (b+\Delta+1,a+\alpha_0 + 1],\\
F_{\eta_1}(e) & \text{ on } e > a + \alpha_0 + 1.
\end{cases}   
$$
Then, because $-\tilde{w} = \Delta - w \in [a+\Delta,b+\Delta]$ and $x_1 = 0, 1$, 
\begin{align*}
P(Y_1= 1,Y_2 = 0|X=x,X_1 = x_1) = & \int F_{\eta_1}(x_1- w)(1-F_{\eta_2}(x-w))f_W(w)dw \\
= & \int F_{\tilde{\eta_1}}(x_1-\tilde{w})(1-F_{\tilde{\eta}_2}(x-\tilde{w}))f_{\tilde{w}}(\tilde{w})d\tilde{w}. 
\end{align*}
Similarly, because $\alpha - \tilde{w} = \alpha_0 - w \in [\alpha_0 + a, \alpha_0 +b]$ and for $e \in (\alpha_0 + a, \alpha_0 +b] \cup (1+\alpha_0 + a, 1+\alpha_0 +b]$, $F_{\tilde{\eta}_1}(e) = F_{\eta_1}(e)$, we have 
\begin{align*}
P(Y_1= 1,Y_2 = 1|X=x,X_1=x_1) = & \int F_{\eta_1}(x_1+\alpha_0 - w)F_{\eta_2}(x-w)f_W(w)dw \\
= & \int F_{\eta_1}(x_1+\alpha - (w+\alpha - \alpha_0))F_{\eta_2}(x-w)f_W(w)dw \\
= & \int F_{\tilde{\eta_1}}(x_1+\alpha - \tilde{w})F_{\tilde{\eta_2}}(x-\tilde{w})f_{\tilde{w}}(\tilde{w})d\tilde{w}. 
\end{align*}
This implies $\alpha_0$ is not identified from the impostor $\alpha$.

\section{Proof of Theorem \ref{thm:twofactor2}}
\label{sec:twofactor2}
\setcounter{equation}{0}
We first show that both $\sigma_0$ and the density of $\eta_2$ are identified. Note $X$ has full support. This implies the density of $V$ denoted as $f_V(\cdot)$ is identified via 
\begin{align*}
f_V(v) = \partial_v E(Y_2|X=v). 
\end{align*}

In addition, we have 
\begin{align*}
f_V(\cdot) = f_{\eta_2}*\phi_{\sigma_0}(\cdot), 
\end{align*}
where $*$ denotes the convolution operator. Suppose $f_{\eta_2}(\cdot)$ and $\sigma_0$ are not identified so that there exist $f_{\eta_2}'(\cdot)$ and $\sigma'$ such that 
\begin{align*}
f_V(\cdot) = f_{\eta_2}'*\phi_{\sigma'}(\cdot).
\end{align*} 
Without loss of generality, we assume $\sigma' \geq \sigma_0$, otherwise, we can just relabel $f_{\eta_2}(\cdot)$ and $f_{\eta_2}'(\cdot)$. Then we have 
\begin{align*}
f_{\eta_2}(\cdot) = f_{\eta_2}'*\phi_{(\sigma^{'2} -\sigma_0^2)}. 
\end{align*} 
By Assumption \textbf{B5}, we have $\sigma^{'} = \sigma_0$, which implies $f_{\eta_2}(\cdot) = f_{\eta_2}'(\cdot)$. 

In the following, we proceed given that $f_{\eta_2}(\cdot)$ and $\sigma_0$ are known. Recall $F_{\eta_1}(\cdot)$ as the CDF of $\eta_1$. Then,  
\begin{align*}
P^{11}(x_1,x) = & P(Y_1=1,Y_2=1|X_1=x_1,X=x)=  \int F_{\eta_1}(x_1+\alpha_0-\gamma_0 w)F_{\eta_2}(x-\sigma_0 w)f_W(w)dw
\end{align*}
and 
\begin{align*}
P^{10}(x_1,x) = & P(Y_1=1,Y_2=0|X_1=x_1,X=x) = \int F_{\eta_1}(x_1-\gamma_0 w)(1-F_{\eta_2}(x-\sigma_0 w))f_W(w)dw.
\end{align*}
Taking derivatives of $P^{11}(x_1,x)$ and $P^{10}(x_1,x)$ w.r.t. $x$, we have 
\begin{align}
\label{eq:p11'}
\partial_x P^{11}(x_1,x) = \int F_{\eta_1}(x_1+\alpha_0-\gamma_0w)f_{\eta_2}(x-\sigma_0 w)f_W(w)dw
\end{align}
and 
\begin{align}
\label{eq:p10'}
-\partial_x P^{10}(x_1,x) = \int F_{\eta_1}(x_1-\gamma_0w)f_{\eta_2}(x-\sigma_0 w)f_W(w)dw.
\end{align}

Applying Fourier transform on both sides of \eqref{eq:p11'} and \eqref{eq:p10'}, we have 
\begin{align}
\label{eq:f11}
\mathcal{F}(\partial_x P^{11}(x_1,\cdot)) = \mathcal{F}_{\sigma_0}( F_{\eta_1}(x_1+\alpha_0-\gamma_0 \cdot)f_W(\cdot))\mathcal{F}(f_{\eta_2}(\cdot))
\end{align}
and
\begin{align}
\label{eq:f10}
\mathcal{F}(-\partial_x P^{10}(x_1,\cdot)) = \mathcal{F}_{\sigma_0}( F_{\eta_1}(x_1-\gamma_0 \cdot)f_W(\cdot))\mathcal{F}(f_{\eta_2}(\cdot)),
\end{align}
where for a generic function $g(w)$, 
\begin{align*}
\mathcal{F}_{\sigma_0}(g(\cdot))(t) = \frac{1}{\sqrt{2\pi}}\int \exp(-2\pi it\sigma_0w)g(w)dw. 
\end{align*}

Then, by \eqref{eq:f11}, we can identify $F_{\eta_1}(x_1+\alpha_0 - \cdot)$ by 
\begin{align*}
F_{\eta_1}(x_1+\alpha_0 - \gamma_0 \cdot) = \mathcal{F}_{\sigma_0}^{-1}\left(\frac{\mathcal{F}(\partial_x P^{11}(x_1,\cdot))}{\mathcal{F}(f_{\eta_2}(\cdot))} \right)(\cdot)/f_W(\cdot).
\end{align*}
Similarly, we can identify 
\begin{align*}
F_{\eta_1}(x_1 - \gamma_0 \cdot) = \mathcal{F}_{\sigma_0}^{-1}\left(\frac{\mathcal{F}(-\partial_x P^{10}(x_1,\cdot))}{\mathcal{F}(f_{\eta_2}(\cdot))} \right)(\cdot)/f_W(\cdot),
\end{align*}
where for a generic function $g(w)$, 
\begin{align*}
\mathcal{F}_{\sigma_0}^{-1}(g(\cdot))(t) = \frac{\sigma_0}{\sqrt{2\pi}} \int \exp(2\pi it\sigma_0w)g(w)dw. 
\end{align*}
By finding the two pairs $((x_1,w), (x_1',w'))$ and $((\tilde{x}_1,\tilde{w}), (\tilde{x}_1',\tilde{w}'))$ such that $w - w' \neq \tilde{w} - \tilde{w}'$, 
\begin{align*}
F_{\eta_1}(x_1+\alpha_0- \gamma_0 w) = F_{\eta_1}(x_1'- \gamma_0 w'), \quad \text{and} \quad  F_{\eta_1}(\tilde{x}_1+\alpha_0- \gamma_0 \tilde{w}) = F_{\eta_1}(\tilde{x}_1'- \gamma_0 \tilde{w}') 
\end{align*}
we can identify both $\alpha_0$ and $\gamma_0$ as the solution of the following linear system: 
\begin{align*}
& \alpha_0+ \gamma_0 (w'-w) = x_1'- x_1 
& \alpha_0 +  \gamma_0 (\tilde{w}'-\tilde{w}) = \tilde{x}_1'- \tilde{x}_1 .
\end{align*}

\section{Nonparametric Factor Structure}\label{semilinear}
\label{sec:nonpar_factor}
\setcounter{equation}{0}
In this section we describe an estimator for the case where we have a nonparametric factor structure.
Recall for this model we had the following relationship between unobservable variables:

\begin{equation}
U=g_0(V)+\tilde \Pi
\end{equation}

where we assumed that $\tilde \Pi \perp V$.

Our goal in this more general setup is to identify and estimate both $\alpha_0$ and $g_0$.
Our identification is based on the condition that

$$x^{}_1 + \alpha_0 - g_0( x^{}) = \tilde{x}^{}_1 - g_0 (\tilde{x}^{}). $$
if and only if 
$$\partial_2 P^{11}(x_1,x)/f_V(x) + \partial_2 P^{10}(\tilde{x}_1,\tilde{x})/f_V(\tilde{x}) = 0.$$

Using the same $i,j$ pair notation as before, this gives us, in the nonparametric case,
\begin{equation}
X_{1i}-X_{1j}=\alpha_0+(g_0(X_i)-g_0(X_j))
\end{equation}
Note the above equation has a ``semi parametric form", loosely related to the model considered in, for example,
\cite{robinson}. However, we point out crucial differences between what we have above and the standard semi linear model.
Here we are trying to identify the intercept $\alpha_0$ which is usually not identified in the semi linear model as it cannot be separately identified from the nonparametric function.
However, note above on the right hand side, we do not just have a nonparametric function of $X_i,X_j$, but the difference of two {\em identical} and {\em additively separable} functions $g_0(\cdot)$.
In fact it is this differencing of these functions which enables us to separately identify $\alpha_0$. Furthermore, as will now see when turning to our estimator of $\alpha_0$,
the structure of the nonparametric component, specifically additive separability of two identical functions of $X_i,X_j$ respectively, can easily be incorporated into our approximation of each of them. From a theoretical perspective separable functions have the advantage of effectively being a one dimensional problem, as there are no interaction terms to have to deal with. It is well known that nonparametric estimation of separable functions do not suffer from the ``curse of dimensionality". See, for example \cite{newey_1994}.
%
%
%
%\subsection{Estimation of the Semilinear Model}

To motivate our estimator of $\alpha_0$ in this nonparametric factor structure model,  we consider modifying methods used to estimate the semi linear model, which is usually expressed as 
\[ y_i=x_i'\beta_0+g(z_i)+\epsilon_i\]
where $y_i$ denotes the observed dependent variable, $x_i,z_i$ are observed regressors, $g(\cdot)$ is an unknown nuisance function, $\epsilon_i$ is an unobserved disturbance term, and $\beta_0$ is the unknown regression coefficient vector which is the parameter of interest. There is a very extensive literature in both econometrics and statistics on estimation and inference methods for this model- see for example \cite{powell-handbook} for some references.

One popular way to estimate this model is to use an expansion of basis functions, for example polynomials or splines to approximate $g(\cdot)$, and from a random sample of
$n$ observations of $(y_i,x_i,z_i)$ regress $y_i$ on $x_i,b(z_i)$ where $b(z_i)$ denotes the set of basis functions used to approximate $g(\cdot)$.
As an illustrative example, assuming $z_i$ were scalar, if one were to use polynomials as basis functions, one would  estimate the approximate model, 
\[ y_i=x_i'\beta_0+\gamma_1z_i+ +\gamma_2 z_i^2+\gamma_3 z_i^3+....\gamma_{k_n}z_i^{k_n} +u_{in}\]
where $k_n$ is a positive integer smaller than the sample size $n$, and $\gamma_1,\gamma_2,...\gamma_{k_n}$ are additional unknown parameters.
This has been done by regressing $y_i$ on $x_i, z_i,z_i^2, ...z_i^{k_n}$, and our estimated coefficient of $x_i$ would be the estimator of $\beta_0$.
The validity of this approach has been shown in, for example, \cite{donaldneweyjmva}.
\noindent
Now for our problem at hand, incorporating a nonparametric factor structure, we propose a kernel weighted least squares estimator.
The weights are as they were before, assigning great weights to pairs of observations where the sum of derivatives of ratios of choice probabilities
are closer to 0.

\noindent
The dependent variable is identical to as before, the set of $n$ choose 2 pairs $X_{1i}-X_{1j}$.
The regressors now reflect the series approximation of $g_0(X_i)-g_0(X_j)$:
\[ g_0(X_i)-g_0(X_j)\approx \gamma_1(X_i-X_j)+\gamma_2(X_i^2-X_j^2)+\gamma_3(X_i^3-X_j^3)+...\gamma_{k_n}(X_i^{k_n}-X_j^{k_n})\]
\noindent
So now our estimator would be to regress $X_{1i}-X_{1j}$ on $1, (X_i-X_j), (X_i^2-X_j^2), ...(X_i^{k_n}-X_j^{k_n})$, using the same weights $\hat \omega_{ij}$ so the estimator of $\alpha_0$, denoted by $\hat \alpha_{NP}$,
would be the coefficient on 1. Specifying the asymptotic properties of this estimator would require additional regularity conditions, notably the rate at which the sequence of integers $k_n$ increases with the sample size $n$.

\noindent
We again only outline these regularity conditions here, and only to establish consistency. 
Since the estimator and proof strategy is very similar to that for the closed form estimator in the online supplement to this paper, here we only state the additional one needed
for the nonparametric model in this section.

\begin{description}
	\item[Assumption BFC] (Basis function conditions) The basis function approximation of the unknown factor structure function satisfies the following conditions:
	\begin{description}
		\item[BFC.1] The number of basis functions, $k_n$, satisfies $k_n\rightarrow \infty$ and $k_n/n\rightarrow 0$.
		\item[BFC.2] For every $k_n$, the smallest eigenvalue of the matrix
		\[E[P_{k_n}P_{k_n}'] \] 
		is bounded away from 0 uniformly in $k_n$, 
		where 
		\[ P_{k_n}\equiv (1, X_i-X_jf, X_i^2-X_j^2, ...X_i^{k_n}-X_j^{k_n})' \]
	\end{description}
\end{description}
\begin{theorem}
	Under Assumptions {\bf I},{\bf K}, {\bf H}, {\bf S}, {\bf PS}, {\bf FK}, {\bf FH}, {\bf BFC},
	\begin{equation}
	\hat \alpha_{NP} \stackrel{p}{\rightarrow} \alpha_0
	\end{equation}
\end{theorem}

\section{Distribution Theory for Closed Form Estimator}
\label{sec:closed_form}
\setcounter{equation}{0}
Many of the basic arguments follow those used in \cite{chenkhanorder} and \cite{chenkhantang}.
% Asymptotics for two step estimator:
Recall what the key identification condition that motivated the weighted least squares estimator:
For  pairs of observations  $(x^{}_1,x^{})$ and $(\tilde{x}^{}_1,\tilde{x}^{})$ in $\Supp(X_1,X)$,	
$$x^{}_1 + \alpha_0 - \gamma_0 x^{} = \tilde{x}^{}_1 - \gamma_0 \tilde{x}^{}. $$
if and only if 
$$\partial_2 P^{11}(x^{}_1,x^{})/f_V(x^{}) + \partial_2 P^{10}(\tilde{x}^{}_1,\tilde{x}^{})/f_V(\tilde{x}^{}) = 0.$$

\noindent
where recall $\partial_2$ denotes the partial derivative with respect to the second argument.
Note that even though the random variable $V$ is unobserved,  the density function $f_V(\cdot)$ above can be recovered from the data 
from the partial derivative of the choice probability in the treatment equation with respect to the regressor in the treatment equation.
Thus the above equation involves the sum of two ratios of derivatives of choice probabilities.

\noindent
Recall  $\theta_0\equiv(\alpha_0,\gamma_0)$.
Our estimator of $\theta_0$ is based on pair of observations from the data set. We will denote the random variables of interest with capital letters, for example $X_i,X_{1i}$,
and realizations of them with lower letters, for example $x_i,x_{1i}$. To denote distinct random variables in the sample when they form pairs, we will use the subscripts $i,j$.

\noindent
Note  from above, we can express the equation where the pairs receive positive weights (those whose derivatives of choice probabilities summed up to 0) as
\begin{equation}
x_{1i}-x_{1j}=\alpha_0+\theta_0(x_i-x_j)
\end{equation}
So this motivates regressing the scalar random variable $x_{1i}-x_{1j}$ on the two by one random vector ${\bf  x}_{ij}\equiv(1,x_i-x_j)$.
We can now see that  if sufficient  such pairs of observations, where the sum of the ratio of derivative of probabilities could be found to equal 0, $\theta_0$ could be recovered as the unique solution to 
the system of equations corresponding to the pairs, as long as the  matrix involving the terms
${\bf  x}_{ij}$ satisfied a full rank condition.
Such an approach is infeasible for two reasons. The first reason is that the probability functions, their derivatives, and hence the ratio of derivatives  are unknown. The second reason is that even if these functions were known, if the probability functions are not discrete valued, 
such ``matches" will occur with probability zero.

\noindent
The first problem can be remedied by replacing the true probability function values with their nonparametric estimates. In the theory here we used a kernel estimator with kernel function $K(\cdot)$ and bandwidth $H_n$, whose properties are discussed below. 
The second problem can be dealt with through the use of ``kernel weights" as has been frequently employed in the semiparametric literature. 

\noindent
Specifically, assuming that the ratio of derivatives of conditional probability functions were known, we use the following weighting function for pairs of observations;
to illustrate let  $P^{k,l,r}$, $k=0,1, l=0,1$ denotes the ratio of derivatives of choice probabilities.
So, for example, $P^{1,1,r}=\partial_2 P^{11}(X_{1},X)/f_V(X) $ 
, where $\partial_2$ denotes the partial derivative with respect to the second argument.
Let $p^{1r}_i,p^{0r}_j$ denote the $i^{th},j^{th}$ realizations of $P^{1,1,r},P^{1,0,r}$ respectively; then
\begin{equation} \label{kernwgt}
\omega_{ij}=\frac{1}{h_{n}}k\left(\frac{p^{1r}_{i}+p^{0r}_{j}}{h_{n}}\right)
\end{equation}
In \eqref{kernwgt} $h_{n}$ is a bandwidth sequence, which converges to zero as the sample sizes increases, ensuring that in the limit, only pairs of observations with probability functions summing up to an arbitrarily  small number receive positive weight.
$k(\cdot)$ is the kernel function, which is  symmetric around 0, and assumed to have compact support, integrate to 1, and satisfy certain smoothness conditions discussed later on.

\noindent
With the weighting matrix defined, a natural estimate of it, $\hat{\omega}_{ij}$ follows from replacing the true 
probability function values with their nonparametric, e.g. kernel, estimates. 
This suggests a weighted least squares estimator of $\theta_0\equiv (\alpha_0,\gamma_0)$, regressing $x_{1i}-x_{1j}$ on
${\bf  x}_{ij}$, with weights $\hat{\omega}_{ij}$.

\noindent
Specifically,  we propose the following two stage procedure. The first stage is the kernel estimator of the ratio of derivatives of probability functions, and the second stage estimator is defined as:
\begin{equation}
\hat{\theta}=(\sum_{i\neq j}\tau_i\tau_j \hat{\omega}_{ij} {\bf  x}_{ij} {\bf  x}_{ij}')^{-1}
(\sum_{i\neq j}-\tau_i\tau_j\hat{\omega}_{ij}{\bf  x}_{ij} \Delta x_{1ij}) 
\end{equation}
where $\Delta x_{1ij} \equiv x_{1i}-x_{1j}$, ${\bf   x}_{ij} \equiv (1, x_i- x_j)$ and $\tau_i\equiv\tau(x_{1i}, x_i)$ is  a trimming function to remove observations where regressors take values near the boundary of its support. 

\noindent
We will outline the asymptotic properties of this estimator. Here we use similar arguments to this used in
\cite{chenkhanorder} and keep our notation as close as possible to that used in that paper. To simplify characterizing the asymptotic properties of this estimator and the regularity conditions we impose,
we first define   the following functions of $P^{k,l,r}$ for $k=l=1, k=1,l=0$ at their $i^{th}$ and $j^{th}$ realized values, denoted by $p^{1r}_i, p^{0r}_j$

\begin{enumerate}
	\item $f_{P^{k,l,r}_0}=f_{P^{k,l,r}_0}(P^{k,l,r}_{0i})$, where $f_{P^{k,l,r}_0}(\cdot)$ denotes the  density function
	of  $P^{k,l,r}_{0i}$.
	\item $\mu_{\tau i}=E\left[\tau_i|P^{k,l,r}_{0i}\right]$ 
	\item $\mu_{\tau x i}= E\left[\tau_i\tilde X_i|P^{k,l,r}_{0i}\right] $
	\item $\mu_{\tau xxi}=E\left[\tau_i\tilde X_i\tilde X_i'|P^{k,l,r}_{0i}\right]$ 
\end{enumerate}

\noindent
$\mu_1(p^{1r}_i,p^{0r}_j)\equiv E[{\bf x}_{ij}{\bf x}_{ij}'|p^{1r}_i,p^{0r}_j]$ where ${\bf x}_i$ denotes the $2\times 1$ vector $(1,x_i)$,  $\mu_0(p^{0r}_j)\equiv E[{\bf x}_j|p^{0r}_j]$,
where ${\bf x}_j$ denotes the $2\times 1$ vector $(1,x_j)$, 
$f_1(\cdot)$ denotes the density function of the random variable $P^{1,1,r}$, $f_0(\cdot)$ denotes the density function of the random variable $P^{1,0,r}$.

\noindent
Our derivation of  the asymptotic properties of this estimator are based on the following assumptions\footnote{For notational convenience here we suppress the presence of the trimming function.}:
\begin{description}
	\item[Assumption I] (Identification) The  $2\times 2$ matrix:
	\[ M_1=E\left[\mu_1(p^{1r}_i,-p^{1r}_i)'f_0(-p^{1r}_i)\right ] \]
	has full rank.
\end{description}

\begin{description}
	\item[Assumption K] (Second stage kernel function) The kernel function $k(\cdot)$ used in the second stage (to match the sum of ratios of derivatives to 0) is assumed to have the following properties:
	\begin{description}
		\item[K.1] $k(\cdot)$ is  twice continuously differentiable, has compact support and integrates to 1.
		\item[K.2] $k(\cdot)$ is symmetric about 0.
		\item[K.3] $k(\cdot)$ is an eighth order kernel:
		\begin{eqnarray*}
			\int u^lk(u)du &=& 0 \ \ \mbox{for } l=1,2,3,4,5,6,7 \\
			\int u^8k(u)du &\neq&  0 
		\end{eqnarray*}
	\end{description}
	\item[Assumption H] (Second stage bandwidth sequence) The bandwidth sequence $h_n$ used in the second stage is of the form:
	\[ h_n=cn^{-\delta} \]
	where $c$ is some constant and $\delta\in(\frac{1}{16},\frac{1}{12})$.
\end{description}

\begin{description}
	\item[Assumption S] (Order of Smoothness of Density and Conditional Expectation Functions)
	\begin{description}
		\item[S.1] The functions $P^{k,l,r}$ are eighth order continuously differentiable with  derivatives that are bounded on the support of $\tau_i$.
		\item[S.2]  The functions $f_{P^{k,l,r}_0}(\cdot)$ (the density function of the random variable $P^{k,l,r}$) and $E[{\bf x}_{i}|P^{k,l,r}=\cdot]$,
		where ${\bf x}_i$ denotes the $2\times 1$ vector $(1,x_i)$
		have order of differentiability of 8, with eight order partial derivatives that are bounded on the support of $\tau_i$. 
	\end{description}
\end{description}

\

\noindent
The final set of assumptions involve restrictions for the first stage kernel estimator of the ratio of derivatives. This involves smoothness conditions on the choice probabilities $P^{k,l,r}_{0i}$, smoothness and moment conditions on the kernel function, and rate conditions 
on the first stage bandwidth sequence.

\

\begin{description}
	\item [Assumption PS] (Order of smoothness of probability functions and regressor density functions) The functions $P^{k,l,r}(\cdot)$
	and $f_{X_1,X}(\cdot.\cdot)$ (the density function of the random vector $(X_1,X)$) are continuously differentiable
	of order $p_2$, where $p_2>{5}$.
	\item [Assumption FK] (First stage kernel function conditions) $K(\cdot)$, used to estimate the choice probabilities and their derivatives is an even function, integrating to 1 and is of order  $ p_2$ satisfying $ p_2>{5} $. 
	\item [Assumption FH] (Rate condition on first stage bandwidth sequence) The first stage  bandwidth sequence $H_n$ is of the form:
	\[ H_n =c_2 n^{-\gamma/k} \]
	where  $c_2$ is some constant and $\gamma$ satisfies:
	\[ \gamma \in \left(\frac{ 2}{ p_2}\left(\frac{1}{3}+\delta\right),\frac{1}{3}-2\delta\right) \]
	where $\delta$ is regulated by Assumption {\bf H}.
\end{description}

\begin{theorem}\label{rootnnn}
	% Let $\tilde f_i$ denote the denote the density function of the regressors used in the first stage choice probability estimation, and let $\tilde f'_i$ denote its derivative.
	%Let $f(\cdot)$ denote the p.d.f. of $\epsilon_i$ and define the following functions of 
	%$P^{k,l,r}_{0i}$:
	%\[ \lambda_{k,l,i}=\frac{t_0^{(2)}/\beta_0^{(1)}\cdot F^{-1}(1-P_{2i}) }{(F^{-1}(1-P_{2i})-F^{-1}(P_{0i}))^2\cdot f\left( F^{-1}(P_{0i})\right)}
	% \]
	
	% \[ {\cal G}_{k,l,i}=E\left[\tilde X_i(X_{1i}-\tilde X_i'\theta_0)|P^{k,l,r}_{i}\right] \]
	%and 
	%\[ \lambda_{2i}=\frac{t_0^{(2)}/\beta_0^{(1)}\cdot F^{-1}(P_{0i})}{(F^{-1}(1-P_{2i})-F^{-1}(P_{0i}))^2\cdot f\left(F^{-1}(1-P_{2i})\right)}
	% \]
	Let
	\begin{equation}
	\psi_i=\psi_{1i}+\psi_{2i}+\psi_{3i}+\psi_{4i}
	\end{equation}
	where $\psi_{ji} \ \ \ j=1-4$ are each mean 0 random variables defined in
	equations \ref{inf1},\ref{inf2},\ref{inf3},\ref{inf4}, respectively,
	%\psi_{1i}=2\tau_if_{P^{k,l,r}_0i}
	%\sum_{k,l=0,1}(y^{k,l}_{i}\tilde f'_i/\tilde f_i-\partial_2P^{k,l,r}_{0i}){\cal G}_{k,l,i}(\mu_{\tau i}\tilde x_{i}-\mu_{\tau x i})
	%\end{equation}
	then under Assumptions {\bf I,K,H,S,PS,FK,FH},
	\begin{equation}
	\sqrt{n}(\hat{\theta}-\theta_0)\Rightarrow N(0,M_1^{-1}V_1 M_1^{-1})
	\end{equation}
	where
	\begin{equation}
	V_1= E[\psi_{i}\psi_{i}']
	\end{equation}
\end{theorem}

\noindent
{\bf Proof:} Let ${\bf x}_{ij}\equiv (1, (x_i-x_j)), \Delta x_{1ij}\equiv x_{1i}-x_{1j} $.
Then we can express:
\[ \hat \theta-\theta_0= \left(\usum \hat w_{ij} {\bf x}_{ij} {\bf x}_{ij}'\right)^{-1} \usum \hat w_{ij}{\bf x}_{ij}(\Delta x_{1ij}-{\bf x}_{ij}'\theta_0) \]
We will first derive a plim for the denominator term and the a linear representation for the numerator.
For the denominator term here we aim to establish that the double sum
$\usum \hat w_{ij} {\bf x}_{ij}{\bf x}_{ij}'$ converges in probability to the $2\times 2$ matrix  $M_1$.
To do so, note by Assumption $K.1$ we can expand $\hat w_{ij}$ around $w_{ij}$. The remainder term involves the difference between the nonparametrically estimated derivative functions and the true derivative functions.
By Assumptions $K,H,S$ this remainder term is uniformly (over the support of the trimming function $\tau(\cdot)$) $o_p(1)$- see e.g. \cite{liparmeter}.
It thus suffices to establish the probability limit of
$\usum w_{ij}{\bf x}_{ij}{\bf x}_{ij}'$. To do so we first wish to determine the functional form of its expectation. 
For notational ease here we let $p^{1r}_i,p^{0r}_j$ denote $i^{th}$ and $j^{th}$ realized values of $P^{1,1,r}, P^{1,0,r}$ respectively, and $\hat p^{1r}_i, \hat p ^{0r}_j$ denote their nonparametric estimators.
Following the same arguments as in \cite{chenkhanorder}, \cite{chenkhantang},
we can write the expectation of $w_{ij}{\bf x}_{ij}{\bf x}_{ij}'$ as 
\[ \int k((p^{1r}_i+p^{0r}_j)/h_n)/h_n \mu_1(p^{1r}_i,p^{0r}_j))f_1(p^{1r}_i)f_0(p^{0r}_j) dp^{1r}_idp^{0r}_j \]
where $\mu_1(p^{1r}_i,p^{0r}_j)\equiv E[{\bf x}_{ij}{\bf x}_{ij}'|p^{1r}_i,p^{0r}_j]$, 
$f_1(\cdot)$ denotes the density function of the random variable $P^{1,1,r}$, $f_0(\cdot)$ denotes the density function of the random variable $P^{1,0,r}$.
Changing variables $u=(p^{1r}_i+p^{0r}_j)/h_n$ and taking limits as $h_n\rightarrow 0$, yields that the above integral is
\[ \int \mu_1(p^{1r}_i,-p^{1r}_i)f_1(p^{1r}_i)f_0(-p^{1r}_i) dp^{1r}_i = E\left[\mu_1(p^{1r}_i,-p^{1r}_i)f_0(-p^{1r}_i)\right ]\]
which is $M_1$. We next turn attention to the numerator term. This term is of the form:
\[  \usum \hat w_{ij}{\bf x}_{ij}(\Delta x_{1ij}-{\bf x}_{ij}'\theta_0) \]
Again, we expand $\hat w_{ij}$ around $w_{ij}$. The lead term in this expansion is of the form:
\[  \usum w_{ij}{\bf x}_{ij}(\Delta x_{1ij}-{\bf x}_{ij}'\theta_0) \]
Note that because $p^{1r}_i+p^{0r}_j=0\Rightarrow \Delta x_{1ij}={\bf x}_{ij}'\theta_0)$ from our identification result, it follows from Assumptions K,H that the lead term is $o_p(n^{-1/2})$.
The linear term in the expansion is of the form
\begin{equation} \label{linearterm0} \usum w_{ij}' ((\hat p^{1r}_i-p^{1r}_i)+(\hat p^{0r}_j-p^{0r}_j)) {\bf x}_{ij}(\Delta x_{1ij}-{\bf x}_{ij}'\theta_0) \end{equation}

We will first focus on the term
\begin{equation} \label{linearterm} \usum w_{ij}' (\hat p^{1r}_i-p^{1r}_i) {\bf x}_{ij}(\Delta x_{1ij}-{\bf x}_{ij}'\theta_0) \end{equation}
Recall $\hat p^{1r}_i$ denotes a ratio of non parametrically estimated terms and $p^{1r}_i$ denotes the ratio of derivatives.
Denote these estimated and true  ratios as $\hat f_{vi} ^{-1} \hat p^1_i$ , $ f_{vi}^{-1} p^1_i$ respectively. Linearizing this ratio, the first term is of the form
$f_{vi}^{-1} (\hat p^1_i-p^1_i)$. So we wish first to evaluate a representation for
\begin{equation} \label{lineartermnum}
\usum w_{ij}' f_{vi}^{-1}(\hat p^{1}_i-p^{1}_i) {\bf x}_{ij}(\Delta x_{1ij}-{\bf x}_{ij}'\theta_0) \end{equation}
\noindent
%Recall $\hat p^1_i$ denotes the ratio of the derivative of a choice probability function to density of the unobserved error term
%In the the treatment equation. Here we treat the latter as known and use the kernel estimator for the former.
Denoting a kernel estimator of the  probability function of the outcome variable as a  function of $\vec x=(x_1,x)$, by
$\hat p(\vec x)= \frac{ \sum_j  y_{1j}K_H(\vec x_j-\vec x)}{\sum_j K_H(\vec x_j-\vec x )}$ where $K(\cdot)$ is our kernel function, $H$ our bandwidth, and $K_H(\cdot)\equiv \frac{1}{H}K(\frac{\cdot}{H})$,
our estimator of the derivative of the probability function is
\[ \hat p^1(\vec x)=\frac{\sum_ky_{1k}K'_H(\vec x_k-\vec x)\frac{1}{H}{\sum_k K_H(\vec x_k-\vec x)}-{\sum_k K'_H(\vec x_k- \vec x)}\frac{1}{H}{ \sum_k  y_{1k}K_H(\vec x_k-\vec x)}}{(\sum_k K_H(\vec x_k-\vec x))^2} \]
We plug in the first of the two terms in the above numerator into \ref{lineartermnum} yielding 
\[ \frac{\usumt w_{ij}'f_{vi}^{-1}(y_{1k}K_H'(\vec x_k-\vec x_i)\frac{1}{H}-p_i^1){\bf x}_{ij}(\Delta x_{1ij}-{\bf x}_{ij}'\theta_0)}{\frac{1}{n}\sum_kK_H(\vec x_k-\vec x_i)} \]
In the above expression, we  replace the denominator term with its plim\footnote{The resulting remainder term, involving the difference between the denominator term and its plim, can shown to be asymptotically negligible, as shown in \cite{chenkhantang}} , which is $f_{\vec X}(x_i)$,
which gives the expression:
\begin{equation} \label{linearb} \usumt \left( \frac{y_{1k}K_H'(\vec x_k-\vec x_i)\frac{1}{H}}{f_{\vec X}(\vec x_i)}-p_i^1\right) f_{vi}^{-1}\Gamma_{ij} \end{equation}
where $\Gamma_{ij}=w_{ij}'{\bf x}_{ij}(\Delta x_{1ij}-{\bf x}_{ij}'\theta_0)$.
Evaluating a linear representation for the above third order $U$ statistic in \ref{linearb}, we first evaluate the expectation of 
$\frac{1}{f_{\vec X}(\vec x_i)}y_{1k}K_H'(\vec x_k-\vec x_i)\frac{1}{H}$ conditioning on $\vec x_i$. This can be expressed after a change of variables  as
\[ \frac{1}{f_{\vec X}(\vec x_i)}\int p(uH+\vec x_i)K'(u)f_{\vec X}(uH+\vec x_i)du\frac{1}{H} \]
Where here $f_{\vec X}(\cdot)$ denotes the density function of $\vec X_i$.
Next we can expand around $uH=0$ inside the integral. The lead term is 0 as $K(\cdot)$ vanishes at the boundary of its support.
The linear term is $p^1(\vec x_i)f_{\vec X}(\vec x_i)+p(\vec x_i)f_{\vec X}'(\vec x_i)$ using that $\int uK'(u)du=-1$.
Thus the conditional expectation of the ratio $\frac{ y_{1k}K'_H(\vec x_k-\vec x_i)\frac{1}{H}}{f_{\vec X}(\vec x_i)}$ is $p^1(\vec x_i)+p(\vec x_i)f_{\vec X}'(\vec x_i)/f_{\vec X}(\vec x_i)$.
The first term, $p^1(\vec x_i)$, cancels out with $p^1(\vec x_i)$ in \ref{linearb}. Now, note the second term in \ref{linearterm},
$\frac{\sum_k K'_H(\vec x_k-\vec x)\frac{1}{H} \sum_k  y_{1k}K_H(\vec x_k-\vec x)}{(\sum_k K_H(\vec x_k-\vec x))^2}$ is by analogous arguments $f_{\vec X}'(\vec x_i)p(\vec x_i)/f_{\vec X}(\vec x_i)+o_p(n^{-1/2})$.
So combining these results one conclusion that can be drawn is an average derivative type result (e.g. \cite{powell_stock_stoker}):
\begin{equation} \label{pss} \avg \hat p^1(\vec x_i)-p^1(\vec x_i)=\avg \left \{ y_{1i}\frac{f_{\vec X}'(\vec x_i)}{f_{\vec X}(\vec x_i)}-p^1(\vec x_i)\right\} +o_p(n^{-1/2}) \end{equation}

So plugging \ref{pss} into \ref{linearb} yields:
\[ \usum \left \{ y_{1i}\frac{f_{\vec X}'(\vec x_i)}{f_{\vec X} (\vec x_i)}-p^1(\vec x_i)\right\}f_{vi}^{-1} \Gamma_{ij}+o_p(n^{-1/2}) \]
As an additional step we want a representation for $\Gamma_{ij}$. By its definition,
\begin{equation} \label{linearc}\usum \Gamma_{ij}=\usum w_{ij}'{\bf x}_{ij} (\Delta x_{1ij}-{\bf x}_{ij}'\theta_0) =\usum \frac{1}{h^2} k'\left(\frac{p^{1r}_i+p^{0r}_j}{h}\right ) \zeta(\vec x_i,\vec x_j) \end{equation}
where $\zeta(\vec x_i,\vec x_j)\equiv {\bf x}_{ij}(\Delta x_{1ij}-{\bf x}_{ij}'\theta_0)$.
To attain this representation, we evaluate the expectation of the term inside the double summation. We express this as 
\[ \frac{1}{h^2} \int k'\left(\frac{p^{1r}_i+p^{0r}_j}{h}\right )\bar \zeta(p^{1r}_i,p^{0r}_j)f_1(p^{1r}_i)f_0(p^{0r}_j) dp^{1r}_i dp^{0r}_j \]
where recall $f_1(\cdot)$ denotes the density function of the random variable $P^{1,1,r}$, $f_0(\cdot)$ denotes the density function of the random variable
$P^{1,0,r}$, and here, $\bar \zeta(p^{1r}_i,p^{0r}_j) \equiv E[\zeta(\vec x_i,\vec x_j)|p^{1r}_i,p^{0r}_j]$ To evaluate the above integral we construct the change of variables
$u=\frac{p_i^{1r}+p_j^{r0}}{h}$ and expand inside the integral. Before expanding the integral is of the form
\[ \frac{1}{h}\int k'(u) \bar \zeta(p_i^{1r},uh-p_i^{1r})f_1(p_i^{1r})f_0(uh-p_i^{1r})du dp_i^{1r} \]
After expanding, the lead term is 0 because the function $k(\cdot)$ vanishes on the boundary of its support.  The next term is of the form:
\[ \int \left(\bar \zeta_2(p_i^{1r},-p_i^{1r})f_1(p_i^{1r}) f_0(-p_i^{1r}) +\zeta(p_i^{1r},-p_i^{1r})f_1(p_i^{1r})f_0'(-p_i^{1r})\right)k'(u)u du dp_i^{1r} \]
From our identification result the above integral simplifies to $-E[\bar \zeta_2(p_i^{1r},-p_i^{1r})f_0(-p_i^{1r})]$ which we will denote by $\Xi_1$.
So plugging this result into \ref{linearterm} we have the following result:
\begin{eqnarray}
\usum f_{vi}^{-1}(\hat p_i^1-p_i^1)\Gamma_{ij}&=&\avg \Xi_1 f_{vi}^{-1} \left \{ y_{1i}\frac{f_{\vec X}'(\vec x_i)}{f_{\vec X}(\vec x_i)}-p^1(\vec x_i)\right\} +o_p(n^{-1/2}) \\
&\equiv & \avg \psi_{1i}+o_p(n^{-1/2} )
\end{eqnarray}
where 
\begin{equation}
\label{inf1}
\psi_{1i}= \Xi_1 f_{vi}^{-1} \left \{ y_{1i}\frac{f_{\vec X}'(\vec x_i)}{f_{\vec X}(\vec x_i)}-p^1(\vec x_i)\right\} 
\end{equation}

\noindent 
We next turn attention to the second term in the linearization of the ratio.
This is of the form :
\begin{equation} \label{linearratio2} \usum \Gamma_{ij} \frac{p_i^1}{f_{vi}^2}  (\hat f_{vi}-f_{vi}) \end{equation}
The term $\hat f_{vi}$ is our kernel estimator of the derivative of the probability function in the  treatment equation:
$\hat f_{vi}=\frac{\partial}{\partial X_i} E[Y_{2i}|X_i]$.  So we can use analogous arguments to attain a linear representation for this $U$-statistic in \eqref{linearratio2} to conclude
\begin{eqnarray}  \usum \Gamma_{ij} \frac{p_i^1}{f_{vi}^2}  (\hat f_{vi}-f_{vi})&=&  \avg \Xi_1f_{vi}^{-2} p_i^1 \left \{ y_{2i}\frac{f_X'(x_i)}{f_X(x_i)}-f_V(x_i)\right\} +o_p(n^{-1/2}) \\
&\equiv & \avg \psi_{2i}+o_p(n^{-1/2} )
\end{eqnarray} 
where 
\begin{equation}
\label{inf2}
\psi_{2i}= \Xi_1 f_{vi}^{-2}p_i^1 \left \{ y_{2i}\frac{f_{ X}'( x_i)}{f_{ X}( x_i)}-f_V( x_i)\right\} 
\end{equation}
Next we can turn attention to the the second term in \eqref{linearterm0}, 
\begin{equation} \label{lineartermb} \usum w_{ij}' (\hat p^{0r}_j-p^{0r}_j) {\bf x}_{ij}(\Delta x_{1ij}-{\bf x}_{ij}'\theta_0) \end{equation}
The term $\hat p^{0r}_j-p^{0r}$ involves the ratio of two derivatives. So we can proceed as before by linearizing this ratio. This will yield the two expressions:
\begin{equation}
\avg \Xi_1 f_{vi}^{-1} \left \{ y_{1i}\frac{f_{\vec X}'(\vec x_i)}{f_{\vec X}(\vec x_i)}-p^0(\vec x_i)\right\} +o_p(n^{-1/2}) 
\equiv  \avg \psi_{3i}+o_p(n^{-1/2} )
\end{equation}
where 
\begin{equation}
\label{inf3}
\psi_{3i}= \Xi_1 f_{vi}^{-1} \left \{ y_{1i}\frac{f_{ \vec X}'( \vec x_i)}{f_{ \vec X}(\vec  x_i)}-p^0( \vec x_i)\right\} 
\end{equation}
and 
\begin{equation}
\avg \Xi_1f_{vi}^{-2} p_i^0 \left \{ y_{2i}\frac{f_X'(x_i)}{f_X(x_i)}-f_V(x_i)\right\} +o_p(n^{-1/2}) 
\equiv  \avg \psi_{4i}+o_p(n^{-1/2} )
\end{equation}
where 
\begin{equation}
\label{inf4}
\psi_{4i}= \Xi_1 f_{vi}^{-2}p_i^0 \left \{ y_{2i}\frac{f_{ X}'( x_i)}{f_{ X}( x_i)}-f_V( x_i)\right\} 
\end{equation}

\noindent
So collecting all results we can conclude that the estimator has the linear representation:
\begin{equation}\label{linearrep}
\hat \theta-\theta_0= M_1^{-1} \avg \psi_i+o_p(n^{-1/2}) 
\end{equation}
where $\psi_i\equiv \psi_{1i}+\psi_{2i}+\psi_{3i}+\psi_{4i}$.

\bibliographystyle{chicago}
\bibliography{overall}

\end{document}